\documentclass[a4paper,11pt]{article}
\pdfoutput=1 

\usepackage{jcappub} 
\usepackage{amsmath,amstext}
\usepackage{hyperref}
\usepackage{aas_macros}
\usepackage{xcolor}
\usepackage{xspace}
\usepackage[caption=false]{subfig}
\hypersetup{colorlinks=true}
\usepackage{natbib}
\newcommand{\planck}{\textsl{Planck}\xspace}
\newcommand{\Quaia}{Quaia\xspace}
\newcommand{\Gaia}{\textsl{Gaia}\xspace}
\newcommand{\unWISE}{\textsl{unWISE}\xspace}
\newcommand{\Catalog}{\Gaia--\unWISE Quasar Catalog\xspace}
\newcommand{\Dz}{\Delta z}
\newcommand{\dz}{\Dz/(1+z)}
\newcommand{\iMpc}{{\rm Mpc}^{-1}}
\newcommand{\nv}{\hat{\bf n}}
\newcommand{\hpix}{{\tt HEALPix}\xspace}
\newcommand{\nmt}{{\tt NaMaster}\xspace}
\newcommand{\tomog}{{\tt Tomographer}\xspace}
\newcommand{\zquaia}{z_\mathrm{Quaia}}

\title{Constraining cosmology with the \Gaia--\unWISE Quasar Catalog and CMB lensing: structure growth}

\author[a,*]{David Alonso}
\note[*]{Corresponding author.}
\affiliation[a]{Department of Physics, University of Oxford, Denys Wilkinson Building, Keble Road, Oxford OX1 3RH, United Kingdom}

\author[b,c,*]{Giulio Fabbian}
\affiliation[b]{Center for Computational Astrophysics, Flatiron Institute, 162 Fifth Avenue, New York, NY, 10010, USA}
\affiliation[c]{School of Physics and Astronomy, Cardiff University, The Parade, Cardiff, Wales CF24 3AA, United Kingdom}

\author[d,*]{Kate Storey-Fisher}
\affiliation[d]{Center for Cosmology and Particle Physics, Department of Physics, New York University, 726 Broadway, New York, NY 10003, USA}

\author[e]{Anna-Christina Eilers}
\affiliation[e]{MIT Kavli Institute for Astrophysics and Space Research, 77 Massachusetts Avenue, Cambridge, 02139, Massachusetts, USA}

\author[a]{Carlos Garc\'ia-Garc\'ia}

\author[b,d,f]{David W. Hogg}
\affiliation[f]{Max Planck Institute for Astronomy, K{\"o}nigstuhl 17, 69117 Heidelberg, Germany}

\author[f]{Hans-Walter Rix}

\emailAdd{david.alonso@physics.ox.ac.uk}
\emailAdd{FabbianG@cardiff.ac.uk}
\emailAdd{k.sf@nyu.edu}

\abstract{We study the angular clustering of \Quaia, a \Gaia- and \unWISE-based catalog of over a million quasars with an exceptionally well-defined selection function. With it, we derive cosmology constraints from the amplitude and growth of structure across cosmic time. We divide the sample into two redshift bins, centered at $z=1.0$ and $z=2.1$, and measure both overdensity auto-correlations and cross-correlations with maps of the Cosmic Microwave Background convergence measured by \planck. From these data, and including a prior from measurements of the baryon acoustic oscillations scale, we place constraints on the amplitude of the matter power spectrum $\sigma_8=0.766\pm 0.034$, and on the matter density parameter $\Omega_m=0.343^{+0.017}_{-0.019}$. These measurements are in reasonable agreement with \planck at the $\sim$ 1.4$\sigma$ level, and are found to be robust with respect to observational and theoretical uncertainties. We find that our slightly lower value of $\sigma_8$ is driven by the higher-redshift sample, which favours a low amplitude of matter fluctuations. We present plausible arguments showing that this could be driven by contamination of the CMB lensing map by high-redshift extragalactic foregrounds, which should also affect other cross-correlations with tracers of large-scale structure beyond $z\sim1.5$. Our constraints are competitive with those from state-of-the-art 3$\times$2-point analyses, but arise from a range of scales and redshifts that is highly complementary to those covered by cosmic shear data and most galaxy clustering samples. This, coupled with the unprecedented combination of volume and redshift precision achieved by \Quaia, allows us to break the usual degeneracy between $\Omega_m$ and $\sigma_8$.}

\begin{document}
\maketitle
\flushbottom

\section{Introduction}\label{sec:intro}
  Much of the progress in constraining the physical parameters governing the initial conditions and evolution of our Universe is currently driven by the analysis of tracers of the large-scale structure. In these analyses, we study the spatial distribution and time evolution of various tracers of the matter density fluctuations, which allows us to constrain the Universe's geometry, as well as the growth of structure within it. Two of the most powerful large-scale structure tracers are weak gravitational lensing and galaxy clustering  \cite{2007.08991,2007.15632,2105.13549}. The former provides largely unbiased maps of the matter fluctuations integrated along the line of sight from the source redshift, while the latter is a biased tracer of these fluctuations at the redshifts of the galaxies being observed.
  
  This complementarity (unbiased vs. biased, cumulative in redshift vs. local) motivates the combined analysis of weak lensing and galaxy clustering data in a technique commonly known as \emph{lensing tomography} (often also labelled ``2$\times$2-point'' analysis) \cite{1507.05551,1805.11525,1909.07412,2004.01139}. In this methodology one measures the two-point auto-correlations of a set of samples of galaxies that are reasonably localised in redshift, together with their cross-correlations with lensing convergence or shear maps from a background source. The cross-correlations effectively ``slice'' the lensing map into its contributions at different redshifts and, when combined with the auto-correlations, enables us to break the degeneracy between galaxy bias and the amplitude and growth of structure. The same technique can be applied to any projected tracer of the large-scale structure other than weak lensing, and can be used to e.g. measure the evolution in the mean gas pressure \cite{1904.13347,1909.09102,2006.14650,2102.07701,2204.01649} and star-formation density \cite{1407.0031,2206.15394}, or to reconstruct the redshift distribution of different probes \cite{1407.7860,1810.00885}.

  The lensing of the Cosmic Microwave Background (CMB) has become a treasure trove for lensing tomography. At $z\sim1100$, the CMB is a lensing source that sits behind all tracers of the large-scale structure, and thus can be used to reconstruct the growth history at all redshifts. Various analyses have made use of lensing tomography to constrain cosmology, using both photometric \cite{1203.4808,1410.4502,2105.03421,2010.00466,2111.09898,2203.12440} and spectroscopic samples \cite{1908.04854,2004.01139,2204.10392}. Similar studies can be carried out in cross-correlation with galaxy weak lensing (cosmic shear), as well as combining all three tracers of structure \cite{2105.12108,2206.10824}. Constraints can be strengthened significantly by adding the auto-correlation of the weak lensing data, in what's commonly called a ``$3\times2$-point'' analysis (sometimes called $5\times2$-point when combining galaxy clustering, cosmic shear, and CMB lensing) \cite{1607.01014,2007.15632,2105.13549,2105.12108,2110.03815,2202.07440,2206.10824,2203.12440,2304.00703,2304.00704}. It is often not possible to fully break the degeneracy between growth and geometry with these datasets, and thus results are often reported in terms of the combination $S_8\equiv\sigma_8(\Omega_m/0.3)^{0.5}$, where $\sigma_8$ is defined as the standard deviation of the linear matter overdensity on spheres of radius $8\,{\rm Mpc}/h$, and $\Omega_m$ is the fractional energy density of non-relativistic matter. A growing number of these analyses have consistently recovered measurements of the amplitude of matter fluctuations that, qualitatively, seem to be in tension with the value inferred by CMB experiments such as \planck \cite{1807.06209}, or the Atacama Cosmology Telescope (ACT) \cite{2007.07288,2304.05203} (the so-called ``$S_8$ tension''). There are hints that the source of this tension may lie in the impact of non-linear baryonic effects, particularly in the case of cosmic shear data, which is more sensitive to small-scale structures ($k\sim0.1-1\,\iMpc$) \cite{2206.11794,2211.11745,2303.05537}. Nevertheless, similar levels of tension have been found from pure CMB lensing tomography analyses \cite{2105.03421,2111.09898}, and thus the origin and actual significance of this potential tension remains unclear.

  In this context, quasars constitute an interesting probe of the large-scale structure, allowing us to trace the matter fluctuations at significantly larger redshifts than standard optical galaxy surveys, covering commensurately large volumes. Their study has been carried out through three-dimensional analysis of their auto-correlation \cite{1705.04718,1904.08859,2201.04679}, as well as their cross-correlation with other tracers, such as the Lyman-$\alpha$ forest \cite{1311.1767}, with the aim of studying both cosmology and the astrophysics of quasars. Since its first detection \cite{1207.4543}, the cross-correlation between quasars and CMB lensing maps has increasingly gained interest, particularly using quasar samples selected from the Sloan Digital Sky Survey \cite{1712.02738,1809.04196,1902.06955}, and as a way to improve our understanding of quasar properties. The use of this cross-correlation for cosmology, however, has so far been limited, although some works have exploited it to constrain the growth history at high redshifts \cite{2105.12108}, as well the level of primordial non-Gaussianity \cite{2305.07650}. The availability of a full-sky quasar sample with a well-understood selection function would facilitate this kind of studies, as well as increase the sensitivity of these measurements by increasing the area overlap with existing CMB lensing maps from experiments such as \planck \cite{1807.06210,2206.07773}.

  In this paper we will study CMB lensing tomography making use of \Quaia \cite{quaia}, a catalog of high-redshift quasars constructed based on the third data release of the \Gaia satellite \cite{2206.05681} as well as \unWISE infrared observations \citep{1008.0031,1405.0308,1909.05444}. Covering the redshift range $z\lesssim4$ with good redshift accuracy ($|\Delta z/(1+z)|<0.01\,(0.1)$ for 62\% (83\%) of the sources), and spanning the full sky, \Quaia allows us to constrain the growth of structure over a range of cosmic times that is highly complementary to those accessed by previous analyses, as well as giving us access to significantly larger scales, where the impact of non-linearities in the matter power spectrum is less important. This complementarity is highly relevant to ascertain whether the source of the $S_8$ tension lies in small-scale physics at low redshifts, an overall lack of power on all scales throughout the Universe's evolution, or a combination of statistical and systematic uncertainties. Furthermore, as we will show, the large-scale reach of the sample will allow us to measure $\sigma_8$ and $\Omega_m$ independently (instead of through the $S_8$ combination).

  The paper is structured as follows. The \Quaia sample and the \planck CMB lensing map used in the analysis are described in Section \ref{sec:data}. Section \ref{sec:th} presents the theoretical framework underpinning CMB lensing tomography. The data analysis methods used are outlined in Section \ref{sec:meth}. Section \ref{sec:res} presents the main results of this analysis, as well as a large battery of tests aimed at confirming their robustness. Our conclusions are then summarised in Section \ref{sec:conc}.

\section{Data}\label{sec:data}
  \subsection{The \Catalog}\label{ssec:data.qso}

    We use the \Catalog, \Quaia \cite{quaia}, as our quasar sample.
    The catalog is all-sky and highly homogeneous, with 1,295,502 sources brighter than its magnitude limit of $G<20.5$ and a median redshift of $z=1.47$.
    It is derived from the quasar candidates identified in the third data release of the space-based \Gaia mission, which classified $\sim$6.6 million sources as potential quasars and estimated their redshifts \Gaia's low-resolution BP/RP spectra.
    \Quaia combines \Gaia data with mid-infrared information from the \unWISE reprocessing of the Wide-field Infrared Survey Explorer (WISE) observations \citep{1008.0031,1405.0308,1909.05444} to improve the sample selection and redshift estimation.
    The redshift estimation uses a $k$-Nearest Neighbors model with \Gaia and \unWISE photometry and the \Gaia-estimated spectral redshift as features, and cross-matched SDSS DR16Q quasars with high-precision spectroscopic redshifts \citep{2007.09001} as labels.
    The \Quaia redshift estimate, $\zquaia$, is taken to be the median of the spectroscopic redshifts of the $K=27$ nearest neighbors, and the uncertainty $\sigma_z$ is their standard deviation.
    Of the \Quaia redshifts, an estimated $62\%$ (83\%) have $|\dz|<0.01$ ($0.1$) with respect to the SDSS redshifts, and 91\% having $|\dz|<0.2$.

    In this analysis we use a modified version of the \Quaia-modelled selection function, re-fit for each of the two redshift bins used for our fiducial results (see Section \ref{ssec:meth.gc}).
    While \Quaia already has relatively minimal selection effects, partly thanks to its space-based observations, the selection function model includes the effects of dust extinction, stellar crowding, and the \Gaia scanning law (as detailed in \cite{quaia}). In Section \ref{sssec:res.rob.internal} we will also explore a brighter sample with $G<20.0$, for which the selection function was also refit.

  \subsection{\planck CMB lensing}\label{ssec:data.planck}
    Our baseline analysis uses the CMB lensing map reconstructed with the Planck PR4 data release based on the NPIPE processing pipeline \cite{2007.04997}. The release improved over various assumptions of the PR3 2018 analysis \cite{1807.06210} that made the analysis sub-optimal in terms of noise treatment and exploited the improved low-level data processing of NPIPE (encompassing additional data and more accurate simulations of the end-to-end data processing)  to reach a $\sim 20\%$ improvement in signal-to-noise at all scales. The details of the analysis are summarized in \cite{2206.07773,2209.07395}. The major pipeline improvements included the adoption of a more optimal generalized minimum-variance estimator (GMV) that performs a joint inverse-variance Wiener filtering of the temperature and polarization CMB maps at the same time instead of treating both observables separately \cite{2101.12193}, as well as an additional post-processing Wiener filtering of the lensing maps according to the local reconstruction noise level on the sky that improves the sensitivity of signal-dominated scales. The PR4 lensing release also includes an estimate of the lensing potential based on the 2018 pipeline and using CMB temperature-only and polarization-only data \cite{1909.02653}. We use these CMB lensing maps to assess the impact of foregrounds in our results, given that the extralagactic foregrounds such as thermal Sunyaev-Zeldovich (tSZ) effect and the Cosmic Infrared Background (CIB) are largely unpolarized at this level of sensitivity. We used the PR3 and PR4 associated lensing simulations to evaluate the consistency between the measurements carried out with the different lensing maps, and to evaluate mask-dependent correction to the lensing map normalization. This is discussed in Section \ref{ssec:meth.kappa}, and in the PR4 lensing documentation\footnote{See \url{https://github.com/carronj/planck_pr4_lensing}.}. All the data products are publicly available through the Planck legacy archive\footnote{\url{http://pla.esac.esa.int/}}. 

\section{Theory}\label{sec:th}
  \subsection{Projected galaxy clustering}\label{ssec:th.gc}
    The first probe we will use in our analysis is the projected distribution of quasars. The core observable in this case is the angular quasar overdensity\footnote{Note that, although normally reserved to label ``galaxies'', we will use the subscript $g$ to denote any quantity related to the quasar samples used.} $\delta_g(\nv)\equiv n_g(\nv)/\bar{n}_g-1$, where $n_g(\nv)$ is the surface density of quasars in the sky direction defined by unit vector $\nv$, and $\bar{n}_g$ is its ensemble average. The angular overdensity is related to the three-dimensional overdensity $\Delta_g({\bf x},t)$ (defined analogously to the angular quasar overdensity), where $\bf x$ is the tracer position, through the line-of-sight (LOS) projection
    \begin{equation}\label{eq:dg_local}
      \delta_g(\nv)=\int d\chi\,H(z)\,p(z)\,\Delta_g(\chi\nv,t(z)),
    \end{equation}
    where $p(z)$ is the redshift distribution of quasars in the sample, $z$ is the redshift to radial comoving distance $\chi$, and $H(z)$ is the expansion rate at redshift $z$\footnote{Note that we use natural units with $c=1$ throughout.}.

    The relation between the quasar overdensity and the underlying matter overdensity $\Delta_m$ is, in general, non-linear, non-local, and stochastic. On sufficiently large-scales ($r\gtrsim10\,{\rm Mpc}$), 
    such as those used in this work, a linear bias relation can be used, so that
    \begin{equation}\label{eq:blinmod}
      \Delta_g=b_g\,\Delta_m+\epsilon,
    \end{equation}
    where $b_g$ is the linear quasar bias and $\epsilon$ is an uncorrelated stochastic field representing the non-deterministic component of the $\Delta_g$--$\Delta_m$ relation. In our fiducial analysis, we will assume that this component is dominated by Poisson shot noise, in which case the power spectrum of $\epsilon$ is simply $P_{\epsilon}(k)=1/\bar{N}_g$, where $\bar{N}_g$ is the mean 3-dimensional quasar density. In general, $\epsilon$ may receive additional contributions due to physics on scales below the Lagrangian size of the protohalo hosting the galaxies composing our samples (e.g. halo exclusion effects). The low density of sources in the quasar sample used here makes it unlikely for these contributions to dominate over the pure Poisson term, but we will study the potential impact of stochastic bias in Section \ref{sssec:res.rob.bz}.

    The observed projected overdensity of sources is also affected by modifications to the properties of the light they emit caused by the large-scale structure. These contributions may be local (e.g. additional red- or blue-shifting caused by peculiar velocities or gravitational potentials), or LOS-integrated (e.g. lensing, integrated Sachs-Wolfe effect, Shapiro time delay) \cite{1105.5280,1105.5292}. Of these, the most relevant effect for distant sources with broad radial projection kernels, such as the samples used here, is the contribution from lensing magnification \cite{1989ApJ...339L..53N,1989A&A...221..221S}. Gravitational lensing causes a coherent distortion in the observed positions of background sources {\sl away} from foreground overdensities, as well as modifies their observed flux, which can add sources to or remove sources from the sample (depending on the slope of the flux distribution, and on the sign of the lensing convergence). The combination of both effects leads to an extra additive contribution to $\delta_g$ in Eq. \ref{eq:dg_local} of the form
    \begin{equation}\label{eq:mag}
      \delta_g^{\rm mag}(\nv)=\int\,d\chi\,(5s-2)\,q_{\rm mag}(\chi)\,\nabla_\perp^2\nabla^{-2}\Delta_m(\chi\nv,t(z)),
    \end{equation}
    where the lensing magnification kernel is
    \begin{equation}\label{eq:qmag}
      q_{\rm mag}(\chi)\equiv\frac{3H_0^2\Omega_m}{2}(1+z)\chi\int_{z(\chi)}^\infty dz'\,p(z')\frac{\chi(z')-\chi}{\chi(z')},
    \end{equation}
    and $s(z)$ is the logarithmic slope of the cumulative number counts of sources at the magnitude limit of the sample:
    \begin{equation}\label{eq:mags}
      s\equiv\left.\frac{d\log_{10}N(<m)}{dm}\right|_{m_{\rm lim}},
    \end{equation}
    where $m$ is the source apparent magnitude, and $m_{\rm lim}$ is the sample's limiting magnitude. Section \ref{sssec:meth.gc.magbias} describes the methods used to calculate $s$ in practice. Finally, $\nabla_\perp^2$ and $\nabla^{-2}$ are the transverse Laplacian and the inverse Laplacian respectively. In the Limber approximation (see Eq. \ref{eq:cllimber}), they partially cancel each other up to a factor
    \begin{equation}
      K_\ell\equiv\frac{\ell(\ell+1)}{(\ell+1/2)^2}=1-\frac{1}{(2\ell+1)^2}.
    \end{equation}
    which can be safely ignored on scales $\ell>10$ (although our theoretical calculations will account for this factor). We will therefore ignore the combination $\nabla_\perp^2\nabla^{-2}$ in what follows for simplicity.

    Finally, it is worth noting that Eq. \ref{eq:mag} assumes that the matter overdensity is related to the Newtonian potential $\Phi$ via Poisson's equation, and that the two scalar metric potentials in the Newtonian gauge, $\Phi$ and $\Psi$ are equal (which is valid for general relativity in the absence of anisotropic stress).

  \subsection{CMB weak lensing}\label{ssec:th.kappa}
    Gravitational lensing perturbs the trajectories of CMB photons, thus distorting the observed pattern of CMB anisotropies, and inducing a statistical coupling between different harmonic scales. This distortion can be used to reconstruct the lensing displacement that causes it, and thus produce maps of the lensing convergence $\kappa(\nv)$, defined as
    \begin{equation}
      \kappa(\nv)=\frac{3H_0^2\Omega_m}{2}\int_0^{\chi_{\rm LSS}}d\chi\,(1+z)\chi\frac{\chi_{\rm LSS}-\chi}{\chi_{\rm LSS}}\,\nabla_\perp^2\nabla^{-2}\Delta_m(\chi\nv,t(z)).
    \end{equation}
    Here, $\chi_{\rm LSS}$ is the comoving distance to the last-scattering surface (LSS). As in the case of lensing magnification, CMB lensing is subject to the same $K_\ell$ correction factor, which can be neglected on the scales used here.

  \subsection{Statistics of projected tracers}\label{ssec:th.cl}
    Consider a projected tracer $u(\nv)$ of a three-dimensional quantity $U({\bf x},z)$:
    \begin{equation}
      u(\nv)\equiv\int d\chi\,q_u(\chi)\,U(\chi\nv,z(\chi)),
    \end{equation}
    where $q_u(\chi)$ is the associated projection kernel. Assuming statistical homogeneity and isotropy, let us define the angular power spectrum $C_\ell^{uv}$, and the three-dimensional power spectrum $P_{UV}(k,z)$ between two pairs of such quantities ($(u,U)$ and $(v,V)$) as:
    \begin{equation}
      \langle u_{\ell m}v^*_{\ell'm'}\rangle\equiv \delta^K_{\ell\ell'}\,\delta^K_{mm'}C_\ell^{uv},
      \hspace{12pt}
      \langle U({\bf k},z)V({\bf k}',z)\rangle\equiv(2\pi)^3\,\delta^D({\bf k}+{\bf k}')\,P_{UV}(k,z).
    \end{equation}
    Here $\langle\cdots\rangle$ denotes ensemble averaging, $\delta^K$ is the Kronecker delta, $\delta^D$ is the Dirac delta, and $u_{\ell m}$ and $U({\bf k},z)$ are the spherical harmonic transform and the Fourier transform, respectively, of the corresponding configuration-space fields:
    \begin{equation}
      u_{\ell m}\equiv\int d\nv\,Y_{\ell m}^*(\nv)\,u(\nv),\hspace{12pt}
      U({\bf k},z)\equiv\int d^3x\,e^{i{\bf k}\cdot{\bf x}}\,U({\bf x},z).
    \end{equation}

    The angular power spectrum can be connected with the three-dimensional $P_{UV}(k)$ via
    \begin{equation}\label{eq:cllimber}
      C_\ell^{uv}=\int \frac{d\chi}{\chi^2}q_u(\chi)q_v(\chi)\,P_{UV}\left(\frac{\ell+1/2}{\chi},z(\chi)\right).
    \end{equation}
    This equation is valid in the Limber approximation \cite{1953ApJ...117..134L,1992ApJ...388..272K}, which is valid when the kernels under study overlap over a range of $\chi$ significantly larger than the correlation length of $U$ and $V$. This is an excellent approximation for the broad CMB lensing kernel, as well as the redshift distribution of the quasar samples we will study.

    Within the linear bias model we use here, both the projected quasar overdensity $\delta_g$ and the lensing convergence $\kappa$ can be treated as tracers of the 3D matter overdensities $\Delta_m$, and therefore $P_{UV}(k,z)$ above is simply the matter power spectrum for all correlations studied. The radial kernels for both tracers are:
    \begin{equation}
      q_g(\chi)=H(z)p(z)b_g+(5s-2)q_{\rm mag}(\chi),
      \hspace{12pt}
      q_\kappa(\chi)=\frac{3H_0^2\Omega_m}{2}(1+z)\chi\frac{\chi_{\rm LSS}-\chi}{\chi_{\rm LSS}},
    \end{equation}
    with the magnification kernel $q_{\rm mag}$ given in Eq. \ref{eq:qmag}.

  \subsection{The importance of high-redshift data}\label{ssec:th.hiz}
    \begin{figure}
      \centering
      \includegraphics[width=0.7\textwidth]{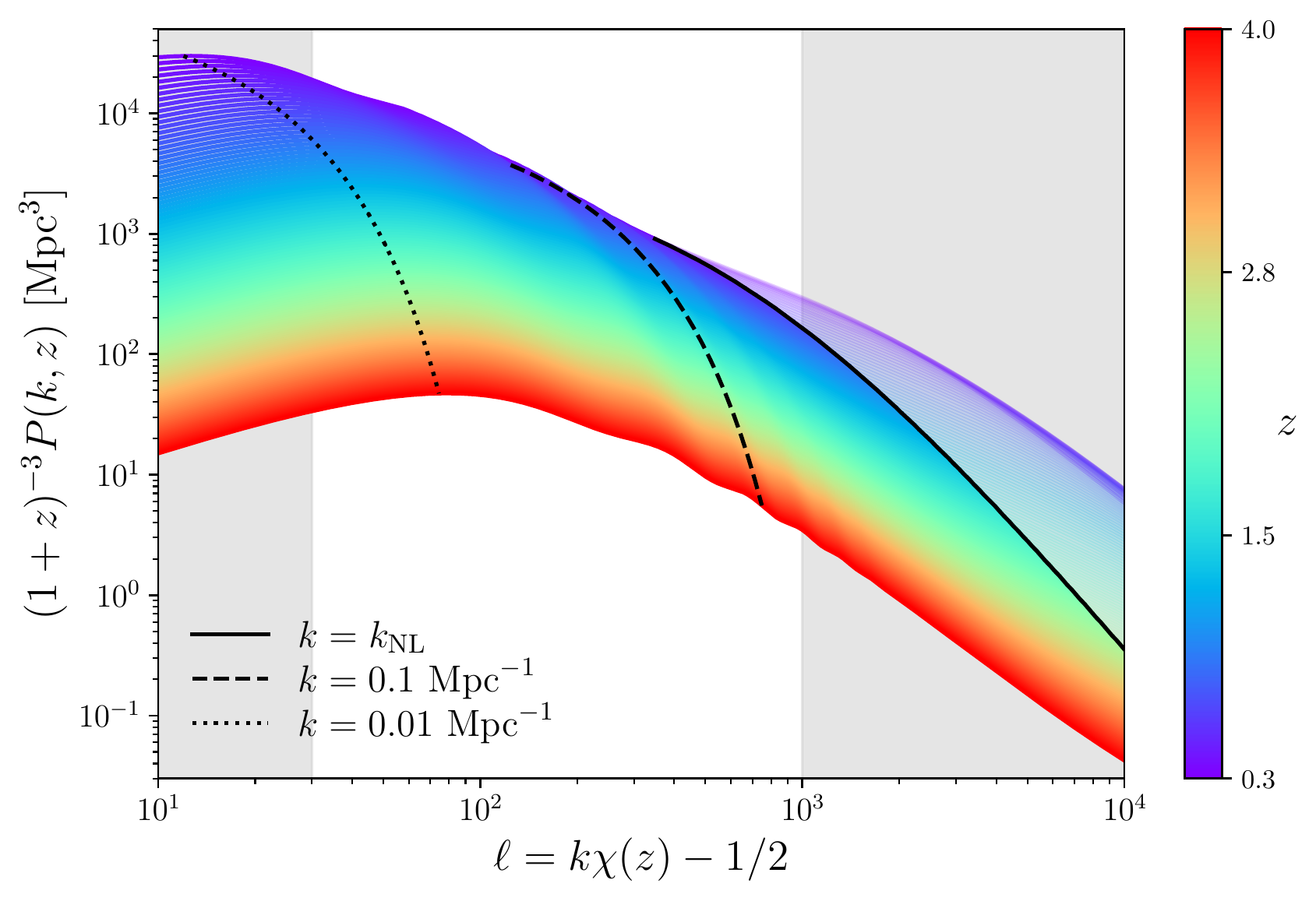}
      \caption{The matter power spectrum as observed by a hypothetical probe sensitive to different redshifts, as a function of the corresponding angular scale $\ell\equiv k\chi-1/2$, where $\chi$ is the comoving distance to that redshift. The central white band roughly corresponds to the range of scales used in cosmic shear analyses. Within these scales, at low redshifts we are mostly sensitive to the monotonic power-law-like part of the power spectrum, which is also affected by non-linear effects (the scale of non-linearities $k_\text{NL}$, defined in Eq. \ref{eq:knl}, is shown by the black solid line). At higher redshifts we are able to better reconstruct the overall shape of the power spectrum at the accessible scales, and to use scales that are less affected by non-linearities. For reference, the plot also shows the position of $k=0.1\,\iMpc$ and $k=0.01\,\iMpc$, roughly corresponding to the location of the BAO wiggles and the horizon scale at matter-radiation equality, respectively.}\label{fig:pk_z}
    \end{figure}

    With the theoretical model described above, we can now explore the benefits of studying the growth of structure at high redshifts, as enabled by the \Quaia quasar sample we use here.

    One of the most compelling reasons to study the history of structure growth, is the ongoing debate surrounding the so-called ``$S_8$ tension'': the fact that a number of low-redshift probes of structure recover an amplitude of the matter fluctuations that is in slight tension with the value inferred from CMB anisotropies. The evidence of this tension from individual experiments, and from some combinations of them, is mild, although not negligible (between 2$\sigma$ and 3$\sigma$), and the fact that several different probes have consistently found low values of the $S_8$ parameter has motivated a widespread search for both further evidence of this tension from other datasets, as well as potential explanations for it in terms of both astrophysical effects and potential new physics.

    The $S_8$ tension is largely dominated by cosmic shear measurements \cite{1906.09262,2007.15633,2105.12108,2105.13543,2304.00701,2306.11124}. The weak lensing kernel for sources at $z\lesssim1$ peaks at $z\sim0.2-0.5$, and extends down to $z\sim0$ \cite{astro-ph/9912508}. Thus, cosmic shear is mostly sensitive to the matter inhomogeneities at low redshifts and, due to projection effects, on relatively small scales $k\sim0.1-1\,\,{\rm Mpc}$. On the other hand, probes such as the CMB lensing convergence auto-correlation, which access significantly higher redshifts (the CMB lensing kernel peaks at $z\sim2$), and are sensitive to larger, more linear scales, find no evidence of a low $S_8$ \cite{2304.05203}. This has motivated the exploration of the mis-modelling of the small-scale matter power spectrum due to non-linearities and baryonic effects as a plausible solution to the $S_8$ tension \cite{2206.11794,2303.05537}. Additional, complementary analyses of structure formation at high redshifts and on linear scales are important in order to investigate this hypothesis; the clustering of high-redshift quasars offers such an opportunity.
    
    Regardless of its potential ability to shed light on the ongoing $S_8$ tension, the study of large-scale structure at redshifts $z\sim2-4$ is interesting for cosmology in its own right. Although we expect the Universe's expansion to be dominated by non-relativistic matter beyond $z\sim1$, direct observations of this epoch are few and far between. The matter fluctuations at $z>2$ could therefore bear the imprint of new early-Universe physics that we are not sensitive to at lower redshift. At the same time, the comoving volume available at high-$z$ gives us access to significantly larger scales than those probed by low-redshift tracers. Fig. \ref{fig:pk_z} illustrates this point. The figure shows the matter power spectrum as it would be observed by an imaginary probe of structure at different redshifts\footnote{We multiply the power spectrum by $1/(1+z)^3$ in order to better distinguish the shape measured at each redshift.}, as a function of the angular multipole, related to the comoving wavenumber via $k\chi\equiv\ell+1/2$, where $\chi$ is the distance to the redshift being probed. The coloured lines show the power spectrum at different redshifts in the range $0.3<z<4$. The solid black line shows the angular multipole corresponding to the non-linear scale $k_{\rm NL}$, defined as the comoving scale at which the overdensity variance reaches unity:
    \begin{equation}\label{eq:knl}
      \sigma^2(<k_{\rm NL}(z))\equiv\int_0^{k_{\rm NL}(z)}\frac{dk\,k^2}{2\pi^2}P_{mm}(k,z)\equiv1.
    \end{equation}
    The power spectra are shown in transparent colours above this scale, which marks the regime in which non-linear and baryonic effects become relevant. The dashed black line shows the position of the wavenumber $k=0.1\,\iMpc$, roughly corresponding to the range of scales where baryon acoustic oscillations are most relevant. The dotted black line shows the wavenumber $k=0.01\,\iMpc$, which corresponds approximately to the horizon scale at the matter-radiation equality, where the power spectrum reaches its maximum. Finally, the shaded gray bands mask out the scales $\ell\lesssim30$, and $\ell\gtrsim1000$ (corresponding to $6^\circ$ and $10'$ respectively), usually discarded in cosmic shear analyses. As we can see, at low redshifts $z\lesssim0.5$, which dominate current cosmic shear data, we mostly have access to the range of scales where the power spectrum is approximately a monotonic power law (projection effects in weak lensing wash out the BAO wiggles), which also receives significant contribution from non-linear scales. In turn, at higher redshifts ($z\gtrsim1$) we gain access to larger scales, and are able to better resolve the shape of the power spectrum, reducing also the contribution from non-linear scales. Measurements at these high redshifts thus can allow us to break degeneracies between different cosmological parameters, which modify the power spectrum in complementary ways.
    \begin{figure}
      \centering
      \includegraphics[width=0.99\textwidth]{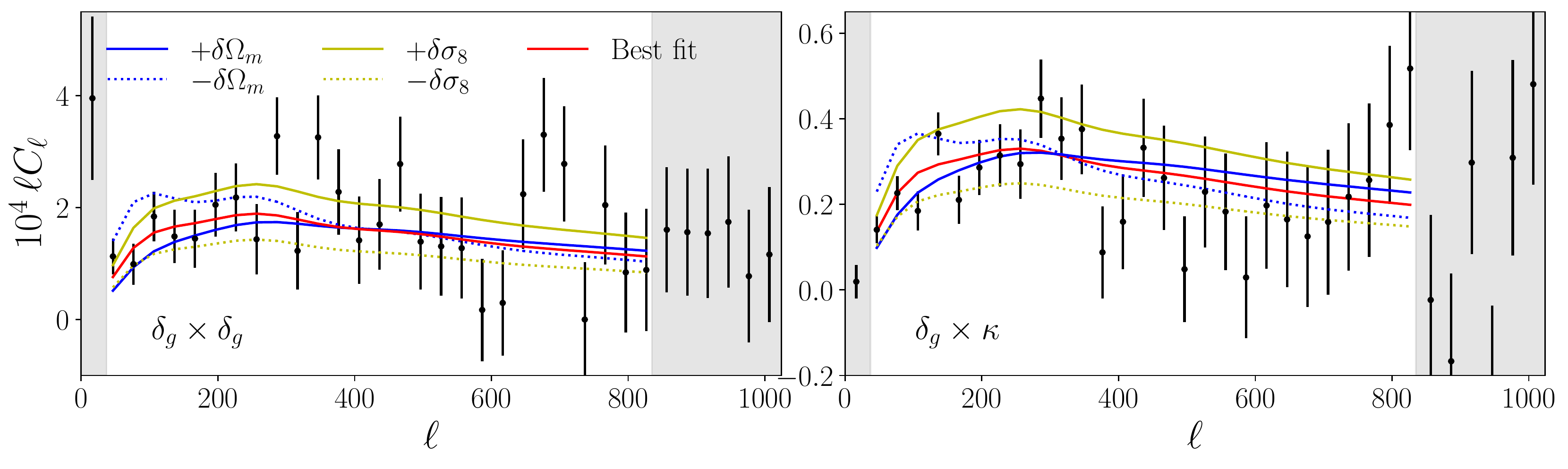}
      \caption{Quasar auto-spectrum (left panel) and cross-spectrum with the CMB lensing convergence (right panel) for the high-redshift bin used in our analysis, centered at redshift $z=2.1$ (see Fig. \ref{fig:dndzs} and Section \ref{ssec:meth.gc}). The measurements are shown as black points with error bars, with the best-fit model shown in red. The blue and yellow lines show the theoretical predictions after changing $\Omega_m$ and $\sigma_8$ by 0.1 in both directions, respectively. On the redshifts and scales our data are sensitive to, the effect of both parameters is markedly different, allowing us to break the degeneracy between them.}\label{fig:clparams}
    \end{figure}

    To illustrate this further, let us briefly jump ahead and examine some of the data we will analyse later on. The left panel of Fig. \ref{fig:clparams} shows the angular auto-correlation of the high-redshift bin we use in our analysis, centered at $z\simeq2$ (see Fig. \ref{fig:dndzs} and Section \ref{ssec:meth.gc}), and the right panel shows the cross-correlation of the same bin with the \planck CMB lensing map. The measured power spectra are shown as black points with error bars, and the red line shows the best-fit model describing these data. The blue solid and dotted lines show the result of changing the cosmic matter fraction $\Omega_m$ by $\delta\Omega_m=\pm0.1$ with respect to its best fit value, while the analogous yellow lines show the result of perturbing $\sigma_8$ by the same amount. We can see that, although on small scales ($\ell\gtrsim400$) both parameters modify the measured power spectra in a similar manner (scaling it up and down), when including the full range of scales used in our analysis (marked by the gray dashed bands), their impact is markedly different. While $\sigma_8$ modifies the overall normalisation of the power spectrum, $\Omega_m$ changes its spectral tilt\footnote{It also changes the amplitude of the BAO peaks through the baryon fraction $f_b\equiv\Omega_b/\Omega_m$, as we keep $\Omega_b$ constant, although we are not sensitive to the BAO wiggles with these data.}. As we will see, this will allow us to break the degeneracy between these parameters, which affects cosmic shear and lower-redshift data sets in general, and measure them independently of one another.

\section{Analysis methods}\label{sec:meth}
  \subsection{Quasar overdensity maps}\label{ssec:meth.gc}
    Our fiducial quasar sample will be the full $G<20.5$ \Quaia catalog, which we divide into two redshift bins ($\zquaia<1.47$ and $\zquaia>1.47$) 
    each containing approximately the same number of sources (647{,}749 and 647{,}753 respectively). Splitting the sample into two bins allows us to explore any trends with redshift (both astrophysical and due to systematics), but we determined that further dividing the sample did not lead to significantly tighter cosmological constraints,  In what follows, we will refer to these two bins as ``Low-$z$'' and ``High-$z$'' respectively. The following subsections describe the methodology used to produce quasar overdensity maps, calibrate their contamination from observational systematics, estimate the slope of the cumulative flux distribution to quantify the impact of magnification bias, and calibrate the redshift distribution of each redshift bin.

    \subsubsection{From catalog to maps}\label{sssec:meth.gc.map}
      We construct a map of the projected quasar overdensity in each redshift bin using the \hpix pixelization scheme \cite{astro-ph/0409513}, with resolution parameter $N_{\rm side}=512$ (corresponding to an angular resolution $\delta\theta\sim0.11^\circ$). In each pixel $p$ we calculate the quasar overdensity as
      \begin{equation}
        \delta_{g,p}=\frac{N_p}{\bar{N}w_p}-1.
      \end{equation}
      Here $N_p$ is the number of quasars in $p$, $\bar{N}$ is the mean number density of quasars in the sample, and $w_p$ is the selection function value in $p$. The selection function is described below in more detail, and can be interpreted as the mean fraction of observed objects (out of those that should have been observed). The mean number density is calculated correcting for the selection function as:
      \begin{equation}\label{eq:nmean}
        \bar{N}=\frac{\sum_pN_p}{\sum_p w_p}.
      \end{equation}

      The method used to construct the selection function is described in detail in \cite{quaia}. In short, a non-linear relation, modelled as a Gaussian process, is found between the number of sources observed in each pixel and the value of a number of systematics templates in those pixels. This relation was found using pixels of resolution $N_{\rm side}=64$ ($\delta\theta\simeq1^\circ$), and considering 4 systematic templates: Milky Way extinction as measured by \cite{1012.4804}, a star map constructed from \Gaia, the specific contribution to the star map from the LMC and SMC, and a measure of depth in \Gaia (the so-called $M_{10}$ quantity \cite{2208.09335}) that maps the \Gaia scanning law and source completeness. It is worth noting that the method intrinsically assumes that the underlying galaxy distribution is statistically isotropic.

      To avoid instabilities when the selection function is close to zero, we mask all pixels where more than half of the sources are missed on average ($w_p<0.5$), and only the resulting unmasked pixels are used in Eq. \ref{eq:nmean}. Elsewhere, the quasar overdensity is set to zero, and the impact of this masking is accounted for via the pseudo-$C_\ell$ estimator (see Section \ref{ssec:meth.cl}). The resulting footprint covers 74\% and 67\% of the sky in the Low-$z$ and High-$z$ bins, respectively. The results presented here were found to be stable against changes in the selection function threshold. In order to downweight regions of the map where a smaller fraction of objects were observed, we again use the selection (truncated at $w_p=0.5$ as above) as a weighting map/mask when estimating power spectra involving $\delta_g$.

      To further remove any residual correlation between the quasar overdensity map and the sky systematics listed above, we make use of linear deprojection, as described in \cite{1609.03577,1809.09603}. Effectively, we fit for, and project out, any linear relation between $\delta_{g,p}$ and the fluctuations of the different systematic templates about their mean\footnote{Note that it is crucially important to subtract the mean from all systematic templates before deprojection since, by design, the overall sky mean is removed when constructing $\delta_{g,p}$.}. Since deprojecting only 4 templates leads to the loss of only a small number of modes, we do not correct for the associated deprojection bias when computing angular power spectra.
      
      The use of linear deprojection after having calibrated the selection function using a Gaussian process (which to some extent should also reconstruct linear trends in the data), may seem redundant. However, it is worth emphasizing that, in the presence of small levels of contamination, where a linear expansion is sufficiently accurate, linear deprojection is exact, non-degenerate with other potential trends in the data, and further guarantees no correlation between the cleaned maps and the different systematic templates. We thus employ it to further reduce the impact of any residual contamination. As we will see in Section \ref{sssec:res.rob.sky}, although we find that deprojection is indeed able to remove the presence of residual systematics in the data, within the scale cuts used in our analysis, their impact is negligible.

    \subsubsection{Magnification bias}\label{sssec:meth.gc.magbias}
      \begin{figure}
        \centering
        \includegraphics[width=0.9\textwidth]{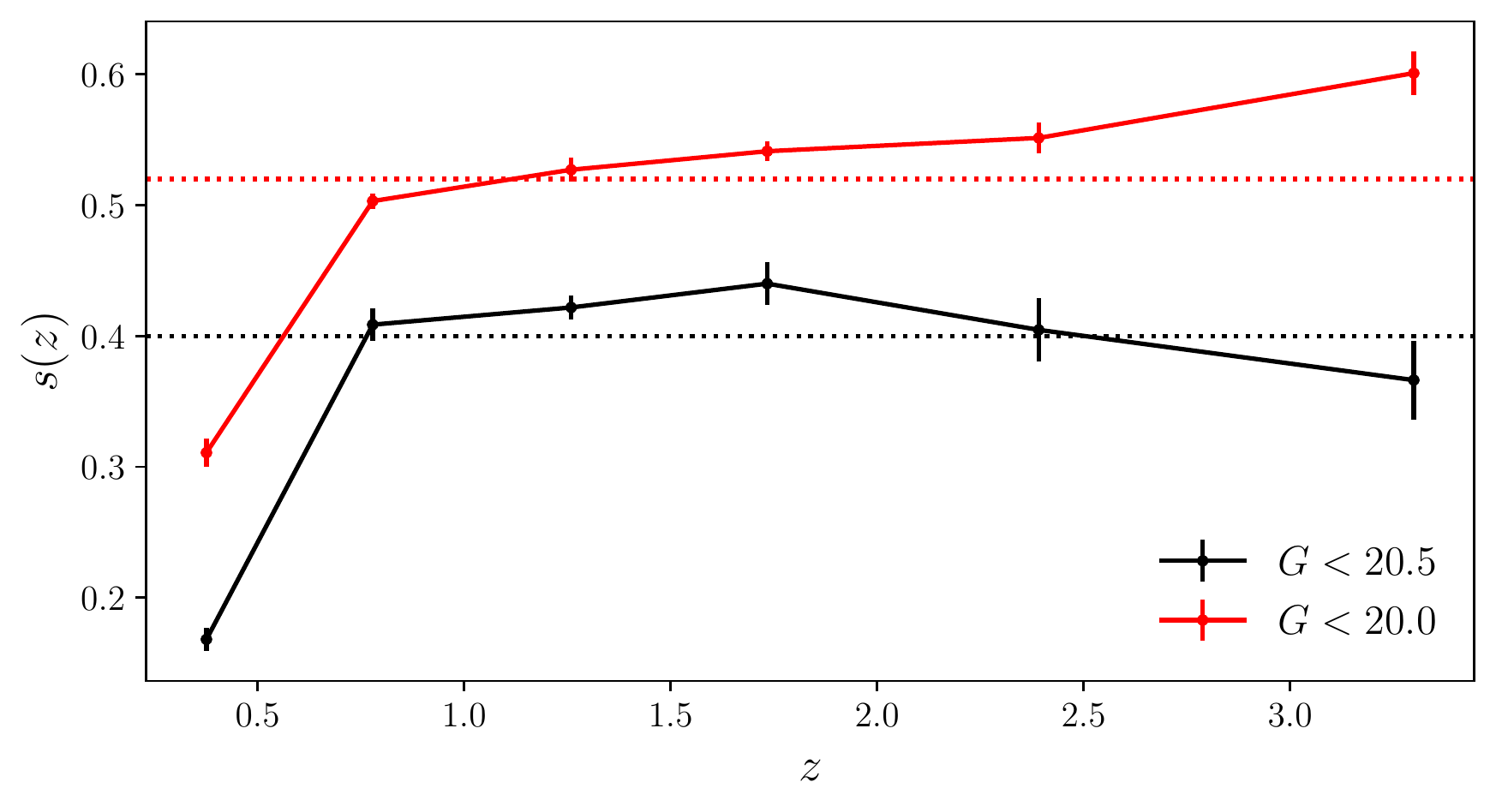}
        \caption{Number counts slope (see Eq. \ref{eq:mags}) governing the impact of lensing magnification, for our fiducial $G<20.5$ sample (black) and for the brighter $G<20$ sample (red), as a function of redshift. The dotted lines show the value of $s$ found for the full sample (i.e. not separated in redshift). The value of $s$ in our fiducial sample, $s\sim0.4$, implies that the impact of lensing magnification on our analysis is negligible.}\label{fig:sofz}
      \end{figure}
      In order to quantify the impact of magnification on the measured power spectra, we need to estimate the slope of the cumulative source number counts with limiting magnitude $s$ (see Eq. \ref{eq:mags}).
      
      The most relevant cuts in the \Quaia sample are in $G$-band magnitude, and in colour space. Magnification leads to a coherent shift in magnitude in all bands, and thus the $G$-band cut is sensitive to it, but not the colour-space cuts. Therefore, as a first method to estimate $s$, we treat \Quaia as a purely magnitude-limited sample, and simply calculate the slope of the number counts at the limiting magnitude\footnote{The raw \Quaia sample extends beyond the magnitude cut $G<20.5$, which allowed us to estimate $s$ via finite differences, comparing the number of sources with $G<20.45$ and $G<20.55$.}. The resulting value of $s$ for the two redshift bins, and for a magnitude limit $G<20.5$ is
      \begin{equation}\label{eq:sfid}
        s(z<1.47)=0.388\pm0.004, \hspace{12pt} s(z\geq1.47)=0.420\pm 0.005.
      \end{equation}
      The statistical uncertainties were estimated via bootstrap resampling. The result for a brighter sample with $G<20$ (which we will explore in Section \ref{sssec:res.rob.internal}) is
      \begin{equation}\label{eq:sG20}
        s(z<1.47,G<20)=0.468\pm 0.005, \hspace{12pt} s(z\geq1.47,G<20)=0.550\pm 0.010.
      \end{equation}

      More in detail, the \Quaia sample is not exactly magnitude-limited, as its definition also involves a cut  in the joint space of magnitude and proper motion. Thus, as an alternative estimate of $s$, we generate two additional catalogs by first perturbing all magnitudes in the raw catalog by $\delta G=\pm 0.05$, and then imposing the cuts defining our sample. $s$ is then estimated from the difference in the number of objects between these two catalogues. The result is:
      \begin{equation}
        s(z<1.47)=0.389 \pm 0.003, \hspace{12pt} s(z\geq1.47)=0.416 \pm 0.004,
      \end{equation}
      in excellent agreement with the more na\"ive estimate above.

      So far we have treated $s$ as a constant in each redshift bin, whereas in reality it likely evolves with redshift. To explore this, we have divided the sample into 6 redshift bins covering the range $\zquaia\in[0,4]$, and calculated $s$ from the number counts slope in each of them. The resulting $s(z)$ for the $G<20.5$ sample is shown in Fig. \ref{fig:sofz}. We see that $s$ only departs from its global value ($s=0.404\pm0.004$ and $s=0.515\pm0.006$ for the $G<20.5$ and $G<20$ samples respectively) significantly at low redshifts, where the contribution from magnification is negligible. We thus assume $s$ to be constant in our fiducial analysis.

      Since, for our sample, the counts slope is close to $s\sim0.4$, the impact of lensing magnification is small, and likely negligible (although we account for it in our theory prediction). This is in contrast with studies carried out with previous optical quasar samples, such as \cite{astro-ph/0502481,astro-ph/0504510}, where lower values were found (e.g. $s=0.2764$ for the DESI-selected quasar sample  \cite{2305.07650}). This may be due to differences in the selection of the various samples, or to the fact that \Quaia targets brighter quasars on average, and hence is sensitive to a different part of the quasar luminosity function. A value of $s<0.4$ corresponds to a negative contribution from magnification, which mostly affects the CMB lensing cross-correlation. To make up for this deficit, we would then infer a larger value of $\sigma_8$, hence the importance of accurately accounting for lensing magnification. In our case, however, we find that this effect can be largely neglected.

      In Section \ref{sssec:res.rob.magbias}, we will present an alternative test, using cross-correlations with a low-redshift sample, to validate our inferred value of $s$ in comparison with the value found for other quasar samples.

    \subsubsection{The redshift distribution}\label{sssec:meth.gc.pz}
      \begin{figure}
        \centering
        \includegraphics[width=0.9\textwidth]{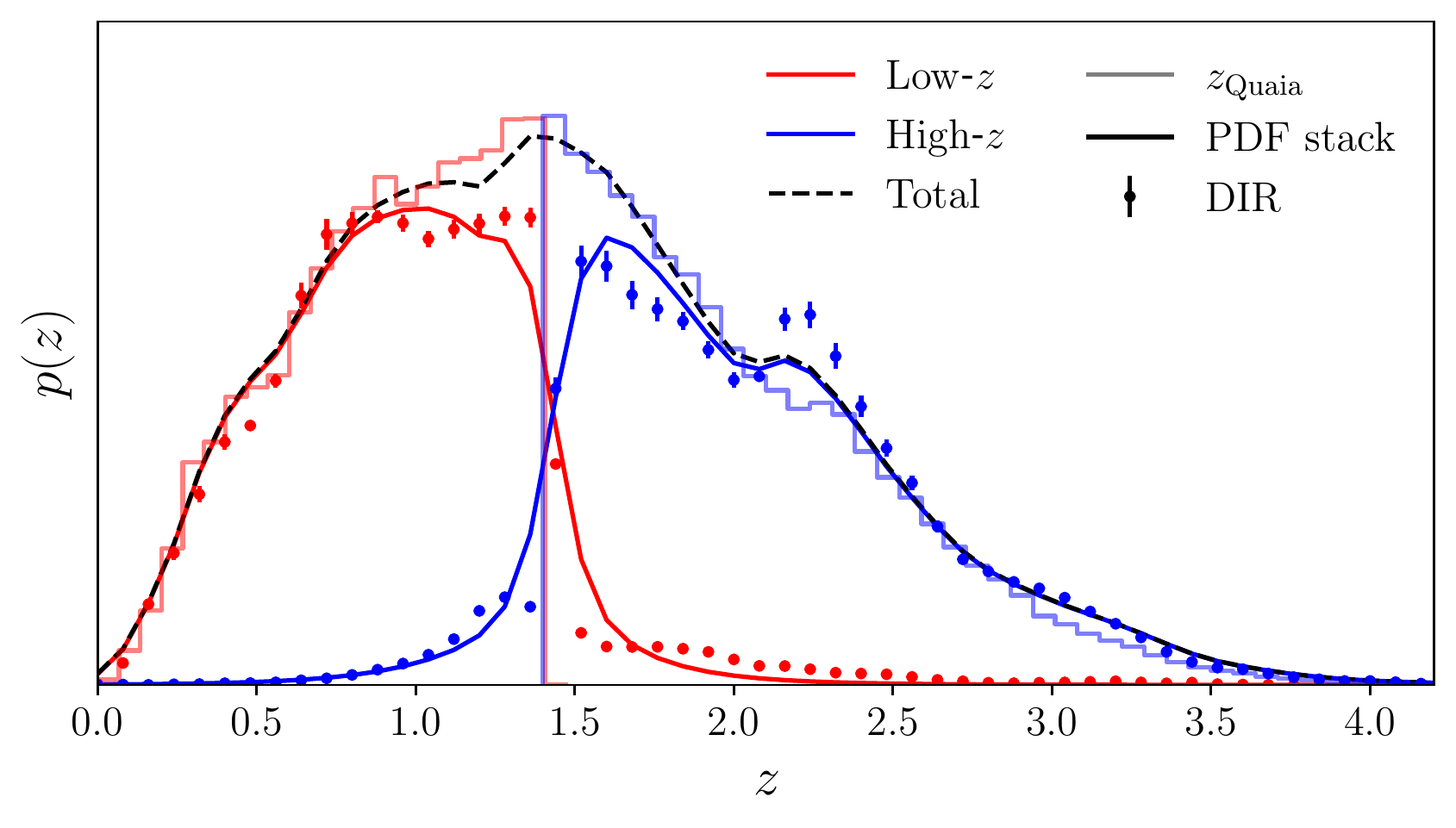}
        \caption{Redshift distribution of the quasar samples studied in this work. The points with error bars show the DIR estimate of the redshift distributions for the two redshift bins used in our fiducial analysis (red and blue respectively). The solid lines show the same redshift distributions estimated via PDF stacking, as described in the main text. The PDF-stacked redshift distribution of the full sample is shown in dashed black (as used in Section \ref{sssec:res.rob.pz}).}\label{fig:dndzs}
      \end{figure}
      The redshift distribution $p(z)$ of the samples under study is a crucial element of the theory prediction, and we must both estimate it and propagate the uncertainties of that estimate to the final inferred cosmological parameters. In this work we make use of two different estimates of the redshift distribution.

      Our fiducial estimate of $p(z)$ is based on the direct calibration (DIR) method of \cite{0801.3822}. In this approach, a cross-matched spectroscopic sample is re-weighted to reproduce the magnitude$/$colour distribution of the target sample. The $p(z)$ is then simply estimated as a weighted histogram of the spectroscopic redshifts. In our case, we make use of the cross-matched SDSS spectroscopic sample used to determine the \Quaia redshift estimate $\zquaia$, and re-weight it in the 5-dimensional space of magnitudes in the $(G,BP,RP,W1,W2)$ bands using a $k$-nearest-neighbours algorithm. This spectroscopic sample contains a total of 243{,}206 sources. The weight of each spectroscopic quasar is proportional to the number of objects in \Quaia within a sphere in magnitude space containing the 10 nearest spectroscopic neighbors (for a more detailed description of the method see e.g. Section 2.2 of \cite{2210.13434}). We verified that the re-weighted spectroscopic catalog follows the same magnitude distribution as \Quaia in all four bands. We estimate the uncertainties in the resulting redshift distribution using jackknife resampling with 100 samples. The estimated $p(z)$s are shown in Fig. \ref{fig:dndzs} as points with error bars for the two redshift bins used in our analysis. We propagate the uncertainties in the DIR redshift distribution making use of the approximate analytical marginalisation method proposed in \cite{2007.14989,2301.11978} (see also \cite{2012.07707}). To account for any potential mis-calibration of the $p(z)$ uncertainties, we verified that our final results did not change significantly when multiplying the jackknife errors by up to a factor of 4.

      One potential drawback of of the DIR-calibrated redshift distributions is the presence of stochastic noise associated with the specifics of the spectroscopic sample used (sample variance, shot noise, selection effects). If this is not correctly captured by the calibrated uncertainties (estimated via jackknife resampling above), this noise may lead to spurious shifts in the final inferred parameters. As an alternative to the DIR $p(z)$, we estimate the redshift distributions of both bins via \emph{PDF stacking}. The \Quaia photometric redshifts $\zquaia$ carry an associated uncertainty $\sigma_z$. Assuming a Gaussian error model, we estimate the sample redshift distribution by adding one Gaussian probability distribution function (PDF) ${\cal N}(\zquaia,\sigma_z^2)$ for each quasar in the sample. The resulting redshift distribution is shown as solid lines in Fig. \ref{fig:dndzs}. We will only use this alternative estimate of the $p(z)$ to quantify the sensitivity of our final results to mis-modelling of the true underlying redshift distribution (which, we will show, is small), and therefore, unlike in our fiducial case, we will not associate any uncertainties to it or marginalise over them.

      As Fig. \ref{fig:dndzs} shows, the redshift distributions exhibit non-negligible tails beyond the threshold $\zquaia$, although both estimates of the $p(z)$ are in qualitative agreement as to their extent. The final cosmological constraints are sensitive to the width of these tails, as they also control the amplitude of the $p(z)$, and hence it is important to ensure that they are correctly modelled. In Section \ref{sssec:res.rob.pz} we will make use of the cross-correlation between quasars in both bins as a sensibility check for the presence and extent of these tails.

    \subsubsection{Bias evolution}\label{sssec:meth.gc.bz}
      Although calibrating the true redshift distribution is important in order to understand the samples under study, the power spectra $(C_\ell^{gg},C_\ell^{g\kappa})$ are in fact only sensitive to the combination $b(z)p(z)$. In other words, the final cosmological constraints depend not only on the redshift distribution, but also on the evolution of the sample's bias with redshift.

      The redshift evolution of the quasar clustering amplitude has been the subject of several works in the literature, including spectroscopic quasars in 2dF \cite{astro-ph/0409314,astro-ph/0607348,1603.04849}, SDSS \cite{0810.4144,0903.3230,1507.08380,1602.09010,1705.04718,1712.02738,1809.04196,1902.06955,2011.01234} and DESI \cite{2304.08427,2305.07650}, and photometrically-selected quasars \cite{astro-ph/0510371,1405.4315,1705.05306,2201.07803}. Although there is significant scatter, especially at high redshifts, the general consensus is that quasar bias evolves strongly with redshift. Thus, although assuming a constant bias within each of the redshift bins studied here would potentially be a more agnostic approach, ignoring the potential evolution within each bin could lead to significant shifts in the final parameter constraints (although in Section \ref{sssec:res.rob.bz}we will show that our final constraints are rather resilient to this effect). Thus, in our fiducial analysis we will assume that the quasar bias within each redshift bin $i$ takes the form
      \begin{equation}\label{eq:bzfid}
        b(z)=b_g^i\,[0.278((1+z)^2-6.565)+2.393],
      \end{equation}
      with $b_g^i$ a free parameter (one in each bin). $b_g^i=1$ would recover the best-fit bias model found by \cite{1705.04718} from the eBOSS quasar sample. To quantify the impact of the assumed level of bias evolution encoded in this model, in Section \ref{sssec:res.rob.bz} we will explore the constant-bias case, and a steeper evolution model, used in \cite{1405.4315}.
      
      Another method to account for the redshift evolution of the sample bias is the so-called \emph{clustering redshifts} approach. This technique \cite{0805.1409,1303.4722,1407.7860,1609.09085} is able to precisely constrain the combination $b(z)p(z)$ from the cross-correlations of the target galaxy sample with a series of spectroscopic (or spectrophotometric) samples with well-determined redshifts. In Section \ref{sssec:res.rob.pz} we will make use of clustering redshifts to validate the robustness of our constraints to uncertainties in the combination $b(z)p(z)$ in a model-independent way.

  \subsection{The CMB $\kappa$ map}\label{ssec:meth.kappa}
    We rotate the harmonic coefficients of the GMV CMB lensing convergence of the PR4 release into Equatorial coordinates, and truncate them to a maximum multipole $\ell_{\rm max}=m_{\rm max}=3N_{\rm side}-1$, before transforming it into a \hpix map of $N_{\rm side}=512$ through an inverse spherical harmonic transform. We use the sky mask provided with the same release, rotate it in real space to Equatorial coordinates, apodize it with a C1 kernel (see Section IIIB of \cite{0903.2350}), and downgrade it to the same angular resolution.

    Compared to previous releases, the PR4 lensing map has a non-trivial normalization that accounts, in a more optimal patch-dependent way, for the effect of anisotropic noise in the CMB maps, as well as its propagation in lensing reconstruction. However, simplifications in the response to the true sky signal and noise can introduce small biases to the normalization, in particular close to the mask boundaries. We compute the multiplicative correction to the estimator normalization (the so-called Monte Carlo (MC) correction) using the simulations associated with the PR4 release (which include realistic sky and noise modelling) measuring the ratio of the power spectrum of the input lensing map spectra to the cross-correlation between the input and the reconstructed lensing maps (see also discussion in \cite{2109.13911,2305.07650}). We compute the power spectra involved in this ratio on the Planck DR4 lensing mask with a binning of $\delta\ell=5$ and interpolate to the relevant effective multipoles of our chosen binning scheme. Such MC correction is of the order of $\sim$5\% on small scales and only becomes significant on the largest scales $\ell\lesssim 10$, which have a minor statistical weight in our analysis. A mismatch in the normalization might also affect the accuracy of the subtraction of the mean-field anisotropy induced by the mask. This mean field becomes important in the CMB lensing autospectrum analysis, and dominates on the largest scales. However, it only contributes to the covariance of the spectrum in the cross-correlations with external tracers and does not introduce any bias. The use of 600 end-to-end Monte-Carlo simulations released with the NPIPE maps to estimate the mean field of the PR4 lensing estimator allowed us to reduce the uncertainty connected to this effect compared to the PR3 release. As such we do not propagate further uncertainties connected to the mean field.

  \subsection{Angular power spectra and covariances}\label{ssec:meth.cl}
    To calculate all auto- and cross-power spectra we make use of the pseudo-$C_\ell$ approach as implemented in \nmt\footnote{\url{https://github.com/LSSTDESC/NaMaster}.} \cite{1809.09603}. The estimator is described in detail in \cite{astro-ph/0105302,1809.09603}, and we only provide a brief description here. The pseudo-$C_\ell$ estimator starts by assuming that all observed maps $\tilde{a}$ are simply given by a locally-weighted version of the true underlying signal maps $a$
    \begin{equation}\label{eq:mask}
      \tilde{a}(\nv)=w_a(\nv)\,a(\nv),
    \end{equation}
    where the weights map $w_a$ is commonly called the ``mask''. In the simplest case, we may consider a binary mask, where $w_a=1$ for observed pixels, and $w_a=0$ for unobserved ones. In practice, the mask can be used as an inverse-variance weight, which down-weights regions of the sky where the uncertainty on the true signal is high. This motivates us using the selection function itself as the mask for the quasar overdensity map, since it tracks the observed number density of the sample which, in the Poisson limit, corresponds to the inverse local variance of $\delta_g$.

    A product in real space (Eq. \ref{eq:mask}) corresponds to a convolution in harmonic space, and the harmonic coefficients of $\tilde{a}$ and $a$ are related to one another via
    \begin{equation}
      \tilde{a}_{\ell m}=\sum_{\ell'm'}W^a_{\ell m,\ell'm'}a_{\ell'm'},\hspace{12pt}
      W^a_{\ell m, \ell'm'}\equiv \int d\nv\,w_a(\nv)\,Y^*_{\ell m}(\nv)\,Y_{\ell'm'}(\nv).
    \end{equation}
    The pseudo-$C_\ell$ of two observed maps $(a,b)$ is defined as
    \begin{equation}
      \tilde{C}^{ab}_\ell\equiv \frac{1}{2\ell+1}\sum_{m=-\ell}^\ell \tilde{a}_{\ell m}\tilde{b}^*_{\ell m}.
    \end{equation}
    It is then straightforward to show that the ensemble average of $\tilde{C}^{ab}_\ell$ is related to the true undelying power spectrum via
    \begin{equation}
      \langle \tilde{C}^{ab}_\ell\rangle = \sum_{\ell'}M^{ab}_{\ell\ell'}C^{ab}_{\ell'}+\tilde{N}^{ab}_\ell,
    \end{equation}
    where $\tilde{N}^{ab}_\ell$ is the pseudo-$C_\ell$ of the noise components of both fields. $M^{ab}_{\ell\ell'}$ is the so-called \emph{mode-coupling matrix}, and encodes the statistical coupling between different harmonic modes caused by the presence of a mask. $M^{ab}_{\ell\ell'}$ depends solely on the pseudo-$C_\ell$ of the masks of both fields, and can be calculated analytically through an ${\cal O}(\ell_{\rm max}^3)$ operation.

    An unbiased estimator for the true power spectrum can thus be constructed by estimating and subtracting the noise bias from $\tilde{C}^{ab}_\ell$, and multiplying the result by the inverse of the mode-coupling matrix. To improve the condition of $M^{ab}_{\ell\ell'}$ before inversion, and to reduce the size of the resulting data vector, it is often common to bin the pseudo-$C_\ell$ into finite bands of $\ell$, commonly called \emph{bandpowers}. The estimated bandpowers $\hat{C}^{ab}_q$ (where $q$ labels each bandpower), can then be related to the true underlying power spectrum via
    \begin{equation}
      \hat{C}^{ab}_q=\sum_\ell {\cal F}^{ab}_{q\ell} C^{ab}_\ell,
    \end{equation}
    where the \emph{bandpower window functions} ${\cal F}^{ab}_{q\ell}$ encode the effects of mode-coupling, binning, and inversion of the binned mode-coupling matrix. We bin all power spectra used here into bandpowers of width $\Delta \ell$=30 between $\ell=2$ and $\ell=3N_{\rm side}-1$.

    Out of all power spectra used in our analysis, only the quasar auto-correlations have a non-zero noise bias due to shot noise. Assuming Poisson statistics, this bias can be estimated analytically as
    \begin{equation}
      \tilde{N}^{gg}_\ell=\frac{\langle w\rangle_p}{\bar{N}_\Omega},
    \end{equation}
    where $\langle w\rangle_p$ is the sky average of the selection function, and $\bar{N}_\Omega\equiv \bar{N}/\Omega_{\rm pix}$ is the angular number density (in units of ${\rm srad}^{-1}$), with the mean number of objects per pixel given by Eq. \ref{eq:nmean}, and $\Omega_{\rm pix}$ the solid angle of each \hpix pixel. In Section \ref{sssec:res.rob.bz} we will explore the impact of residual non-Poissonian shot noise in the data.

    All power spectra are corrected for the effects of the pixel window function as described in \cite{1809.09603}, although we find the impact of this correction to be negligible on the scales used here. Besides this, as described in the previous section, the quasar-$\kappa$ cross-correlation is corrected for the CMB lensing transfer function. Finally, as discussed in Section \ref{sssec:meth.gc.map}, we make use of linear deprojection as implemented in \nmt, ignoring the negligible deprojection bias it incurs in our case.

    We calculate the covariance matrix of the estimated power spectra analytically, following the methodology described in \cite{1906.11765}. The method assumes that all fields involved are Gaussian, which is sufficiently accurate in our case, since all fields under study are largely noise-dominated. The approach described in \cite{1906.11765} has been thoroughly validated, and accurately accounts for the effects of mode-coupling caused by the mask. To achieve this, the method only assumes that the true input power spectra are smooth functions of $\ell$ that do not vary significantly compared to the width of the mode-coupling kernel caused by the masks involved (the so-called \emph{Narrow-Kernel Approximation}, NKA). Given the broad sky coverage of the probes we study, this approximation is only mildly inaccurate on the largest scales, which we discard in our analyses in any case.

  \subsection{Likelihood and priors}\label{ssec:meth.like}
    To infer free parameters of our model $\vec{\theta}$, given our data ${\bf d}$, we make use of a Gaussian (multivariate normal) likelihood:
    \begin{equation}\label{eq:likeG}
      -2\log p({\bf d}|\vec{\theta})=({\bf d}-{\bf t}(\vec{\theta}))^T{\sf C}^{-1}({\bf d}-{\bf t}(\vec{\theta}))+K,
    \end{equation}
    where ${\bf t}(\vec{\theta})$ is the theoretical prediction for ${\bf d}$, ${\sf C}$ is the covariance matrix of ${\bf d}$, and $K$ is an arbitrary normalisation constant that does not depend on $\vec{\theta}$. The \emph{posterior} distribution $p(\vec{\theta}|{\bf d})$ is calculated from the likelihood above using Bayes' theorem
    \begin{equation}
      \log p(\vec{\theta}|{\bf d})=\log p({\bf d}|\vec{\theta})+\log p(\vec{\theta})+K',
    \end{equation}
    where $p(\vec{\theta})$ is the \emph{prior}.

    \begin{table}
        \centering
        \begin{tabular}{|c|l||c|l|}
           \hline
           Parameter & Prior & Parameter & Prior \\
           \hline
           $\sigma_8$ & $U(0.5,1.2)$ & $b_g^i$ & $U(0.1,3.0)$ \\
           $\Omega_m$ & $U(0.05, 0.7)+{\rm BAO}$ & $A^i_{\rm SN}(^*)$ & $N(1, 0.1^2)$ \\
           $h$ & $U(0.4,1.0)+{\rm BAO}$ & $\Omega_mh^3(^\dagger)$ & $N(0.09633, 0.00029^2)$ \\
           \hline
        \end{tabular}
        \caption{Free parameters and priors used in our model. We use BAO data from BOSS and eBOSS \cite{1607.03155,2007.08993} to put a joint prior on $\Omega_m$ and $h$. $(^*)$The shot noise amplitudes $A_{\rm SN}^i$ are fixed to their Poisson level in our fiducial analysis, and we vary them with a 10\% prior to test the dependence of our results on this assumption. $(^\dagger)$In Section \ref{sssec:res.rob.prior} we explore the impact of imposing a prior on the horizon scale at recombination from the positions of the acoustic peaks in the CMB power spectrum. Effectively, this puts a tight constraint on the combination $\Omega_mh^3$ as shown on the table.}
        \label{tab:priors}
    \end{table}
    In our case, ${\bf d}$ is a collection of power spectra, comprising the quasar auto-correlations of the redshift bins under study, and their cross-correlations with the CMB lensing convergence. ${\sf C}$ is the covariance matrix of these measurement, calculated as described in the previous section, and the model used to calculate the theoretical prediction ${\bf t}$ was presented in Section \ref{sec:th}. We only include bandpowers in ${\bf d}$ satisfying the following scale cuts:
    \begin{itemize}
      \item We discard the first bandpower ($\ell<32$) in all power spectra, where the impact of systematic contamination was found to be most relevant (see Section \ref{sssec:res.rob.sky}).
      \item We impose a high-$\ell$ cut at $\ell_{\rm max}=k_{\rm max}\chi(\bar{z})$, where $\bar{z}$ is the mean redshift of the bin under study, and $k_{\rm max}$ is the 3D wavenumber parametrizing our small-scale cut. In our fiducial results we choose $k_{\rm max}=0.15\,\iMpc$, corresponding to $\ell_{\rm max}=(497,\:827)$ in the two redshift bins under study. The main motivation for this cut is ensuring the validity of the linear bias model used in our theoretical predictions. We will study the impact of this choice in Section \ref{sssec:res.rob.internal}).
    \end{itemize}
    With these scale cuts, we are left with $16$ and $27$ bandpowers for power spectra involving the Low-$z$ and High-$z$ quasar redshift bins, respectively, and thus the fiducial size of the data vector combining all auto- and cross-correlations is $N_d=86$. It is worth noting that the Gaussian likelihood of Eq. \ref{eq:likeG}, with a parameter-independent covariance, has been found to be an excellent approximation to the true likelihood of power spectra on scales $\ell\gtrsim10$ \cite{0801.0554,1811.11584}.

    \begin{figure}
      \centering
      \includegraphics[width=0.8\textwidth]{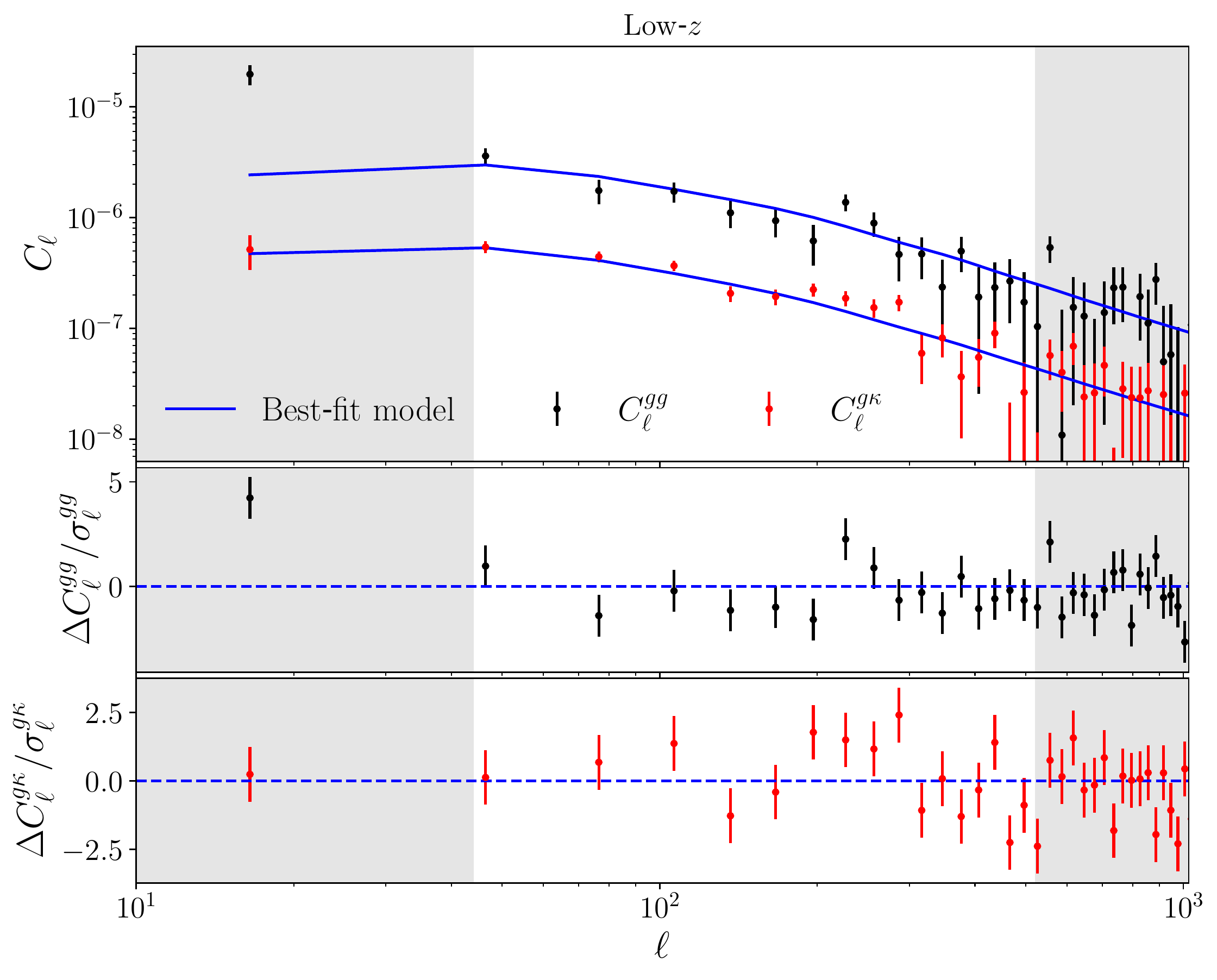}
      \includegraphics[width=0.8\textwidth]{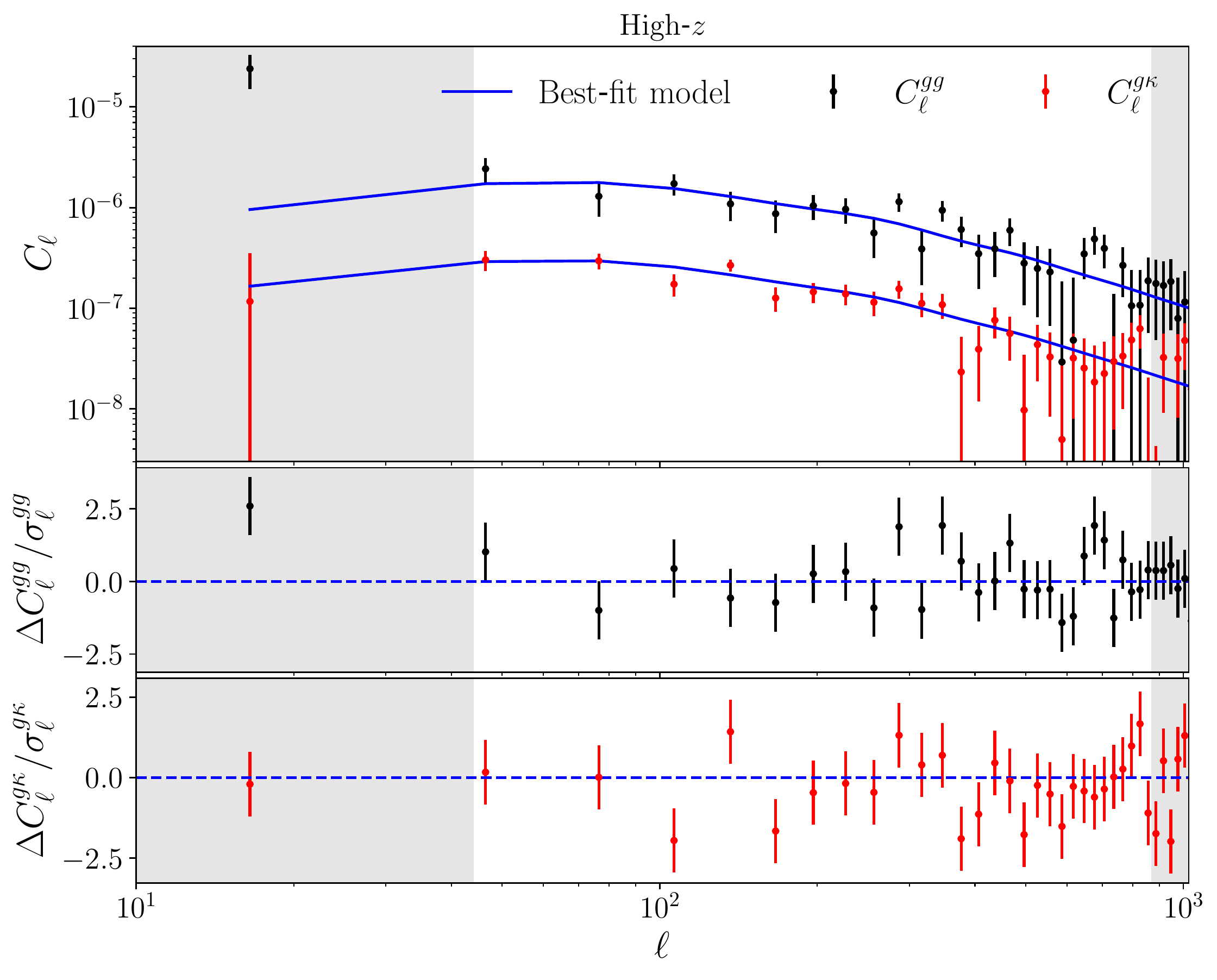}
      \caption{Angular auto-power spectra of the quasar overdensity field (black) and cross-correlation with the \planck CMB lensing convergence map (red) in the two \Quaia redshift bins (top and bottom panels for the Low-$z$ and High-$z$ bins respectively). The blue lines show the best-fit theoretical prediction found in our fiducial configuration. The bottom sub-plots in each panel show the residuals with respect to the best-fit prediction normalised by the $1\sigma$ uncertainties. The central white band shows the angular scales used in our analysis for each redshift bin.}\label{fig:cls}
    \end{figure}

    Within the $\Lambda$CDM model, the main cosmological parameters that affect the power spectra considered here are the total non-relativistic matter density parameter $\Omega_m$, the amplitude of linear matter fluctuations $\sigma_8$, and the expansion rate $h$, and we will vary all three of them in our likelihood. We will fix the baryon density parameter, and the scalar spectral index to the best-fit values found by \planck: $\Omega_bh^2=0.0224$, $n_s=0.9665$ \cite{1807.06209}. Our data is not able to fully break the degeneracy between $\Omega_m$ and $h$ and hence we will impose a prior on both background parameters using low-redshift baryon acoustic oscillations (BAO) data. In particular, we include three measurements of the angular and radial BAO scales from Luminous Red Galaxies (LRGs) at redshifts $z=0.38,\,0.51,\,0.698$ obtained by the BOSS and eBOSS surveys in the DR12 and DR16 of the Sloan Digital Sky Survey \cite{1607.03155,2007.08993}. In Section \ref{sssec:res.rob.prior} we will explore the sensitivity of our final constraints to this choice of prior.

    Besides the cosmological parameters, we consider the two linear bias parameters $\{b_g^1,b_g^2\}$ with flat priors in the range $b^i_g\in[0.1,3.0]$. We assume that the quasar bias evolves as in Eq. \ref{eq:bzfid} in our fiducial analysis, and explore the impact of this assumption on our final results in Section \ref{sssec:res.rob.bz}. We will also quantify the impact of non-Poissonian shot noise on the final parameter constraints by freeing-up the amplitude of the shot-noise contribution to the quasar auto-correlations. This will add two new free parameters $(A_{\rm SN}^1,\,A_{\rm SN}^2)$, one for each redshift bin, with a 10$\%$ Gaussian prior centered around 1. Since both parameters enter linearly in the theoretical prediction, we will marginalise over them analytically. Table \ref{tab:priors} lists all the parameters of the model, and their priors.

    As described in Section \ref{sssec:meth.gc.pz} we marginalise over the uncertainties in the calibrated redshift distribution. We do so following the analytical approach of \cite{2007.14989}, which results in a modification of the power spectrum covariance matrix, without adding any new free parameters to the model.

    All theoretical calculations were carried out using the Core Cosmology Library\footnote{\url{https://github.com/LSSTDESC/CCL}} ({\tt CCL} \cite{1812.05995}). The linear matter power spectrum was computed with CLASS \cite{1104.2933}, and we used the HALOFit prescription of \cite{1208.2701} to connect it with the non-linear power spectrum. We sample the posterior distribution using a the Metropolis-Hastings Monte-Carlo Markov Chain (MCMC) algorithm as implemented in {\tt Cobaya} \cite{2005.05290}. We determine chains to be converged when the Gelman-Rubin statistic reaches $R<0.01$. We will present parameter constraints by quoting the peak of the marginalised 1D distribution, together with its $68\%$ asymmetric confidence-level interval.

\section{Results}\label{sec:res}
  \subsection{Fiducial cosmological constraints}\label{ssec:res.fid}
    \begin{figure}
      \centering
      \includegraphics[width=0.9\textwidth]{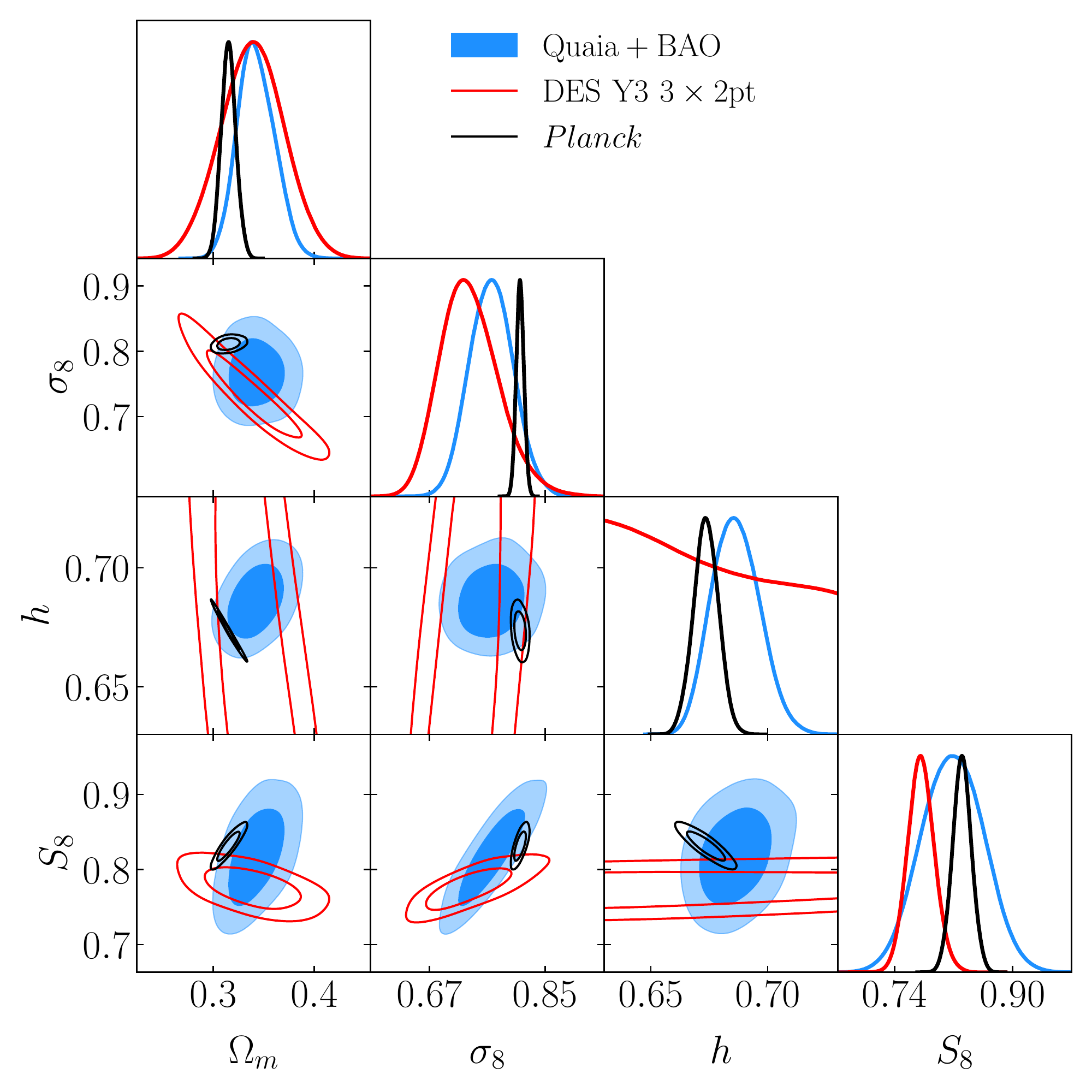}
      \caption{Cosmological constraints from our fiducial analysis of \Quaia with \planck CMB lensing, including a BAO prior from SDSS (filled blue contours), compared with those found by \planck (black), and the Y3 DES 3$\times$2pt analysis (red).}\label{fig:triangle_main}
    \end{figure}
    \begin{figure}
      \centering
      \includegraphics[width=0.99\textwidth]{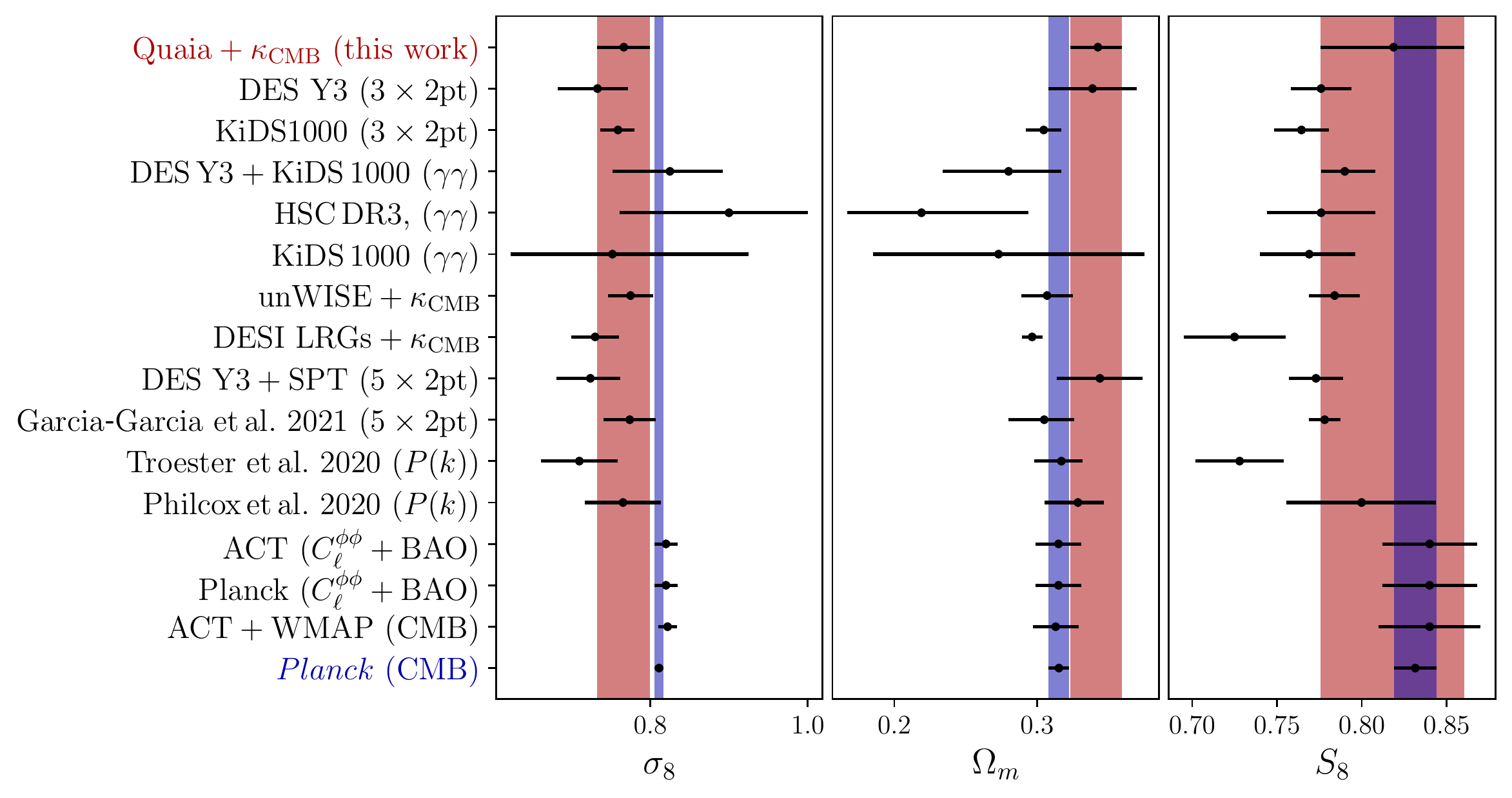}
      \caption{Visual representation of the 68\% confidence level constraints on $\sigma_8$, $\Omega_m$, $S_8$ found in this analysis (salmon band), and found in other datasets. The points shown correspond to cosmic-shear analyses (marked $\gamma\gamma$ \cite{2007.15633,2105.13543,2304.00701}), $3\times2$pt and $5\times2$pt data \cite{2007.15632,2105.13549,2105.12108,2206.10824}, lensing tomography \cite{2105.03421,2111.09898}, full-shape power spectrum analyses \cite{1909.11006,2002.04035}, the CMB lensing power spectrum $+$ BAO data \cite{2206.07773,2304.05203}, and CMB data \cite{1807.06209,2007.07288}. The \planck constraints are shown as a blue vertical band.}\label{fig:results_experiments}
   \end{figure}
    In our fiducial setup, we consider the $G<20.5$ \Quaia sample split into two redshift bins (see Fig. \ref{fig:dndzs}), with DIR-calibrated redshift distributions, fixed magnification bias slope parameters as in Eq. \ref{eq:sfid}, assuming a linear bias evolving as in Eq. \ref{eq:bzfid}, and purely Poisson shot noise. We use only scales with $k<k_{\rm max}=0.15\,\iMpc$, and impose a BAO prior on $(\Omega_m,h)$. 

    Fig. \ref{fig:cls} shows the angular power spectra measured in both bins ($C_\ell^{gg}$ in black, $C_\ell^{g\kappa}$ in red), together with the best-fit theoretical prediction (blue lines). The shaded bands show the angular scales discarded in our analysis. The bottom panels of both figures show the residuals with respect to the best-fit prediction normalised by the $1\sigma$ uncertainties. We find that the best-fit model provides a reasonable fit to the data, with a $\chi^2=96.9$. With $N_d=86$ data points and $N_p=5$ free parameters, the associated probability to exceed (PTE) is $p=0.11$, which corresponds to an adequate fit.

    The cross-correlation between \Quaia and the \planck CMB lensing map is detected, within the scales used in this analysis, with a signal-to-noise ratio of $16.9$ and $13.6$ in the Low-$z$ and High-$z$ bins respectively, for a total detection significance of $21.7\sigma$. This is actually stronger than the significance of the clustering auto-correlation signal itself ($9\sigma$ and $8.4\sigma$ respectively), due to the impact of shot noise. Hence the cross-correlation carries a larger statistical weight in our final cosmological constraints.

    The constraints on the cosmological parameters $(\sigma_8,\Omega_m,h,S_8)$ found with this fiducial setup are shown in blue in Fig. \ref{fig:triangle_main}. As explained in Section \ref{ssec:th.hiz}, we are able to break the degeneracy between $\sigma_8$ and $\Omega_m$, measuring them independently:
    \begin{equation}
      \sigma_8=0.766\pm 0.034,\hspace{12pt}\Omega_m=0.343^{+0.017}_{-0.019}.
    \end{equation}
    The figure also shows, in red and black, the constraints on the same parameters found by the DES Y3 3$\times$2-point analysis \cite{2105.13549}, combining cosmic shear and galaxy clustering from the MagLim sample, and by \planck \cite{1807.06209}, respectively. Our results are in reasonable agreement with the \planck measurements, with only a mild tension at the $1.3\sigma$ and $1.4\sigma$-level for $\sigma_8$ and $\Omega_m$ respectively. The constraints are also fully compatible with those found by DES Y3.

    It is worth placing our constraints in the context of measurements obtained by other experiments. Fig. \ref{fig:results_experiments} shows our constraints on $\sigma_8$, $\Omega_m$, and $S_8$, together with those found in some of the latest analyses of galaxy clustering, cosmic shear, and CMB data. Our constraints are in broad agreement with those found by most of these experiments. We find that we are able to place a competitive constraint on $\sigma_8$, unmatched by cosmic-shear experiments due to its degeneracy with $\Omega_m$, and comparable to that found by $3\times2$pt, $5\times2$pt, and lensing tomography analyses, although with a significantly smaller sample size in the case of \Quaia. Our constraint on $S_8$, in turn, is less stringent than weak lensing and $3\times2$pt analyses, but comparable with those found by full-shape power spectrum analyses in spectroscopic galaxy surveys \cite{1909.11006,2002.04035}.
    
    As highlighted in Section \ref{ssec:th.hiz}, this is one of the highest-redshift constraints on the amplitude of matter fluctuations. That it agrees reasonably with probes at both higher and lower redshifts is therefore a non-trivial result that confirms the ability of the simple $\Lambda$CDM model to describe the large-scale properties of the Universe across most of its history.

  \subsection{Robustness tests}\label{ssec:res.rob}
  \begin{table}
    \begin{center}
      \begin{tabular}{|l|lll|}
        \hline\hline
        Case & $\sigma_8$ & $\Omega_m$ & $S_8$ \\[0.5ex]
        \hline
        1. ${\bf Fiducial}$ & $0.766\pm 0.034$ & $0.343^{+0.017}_{-0.019}$ & $0.819\pm 0.042$\\[0.5ex]
        2. $k_{\rm max}=0.1\,{\rm Mpc}^{-1}$ & $0.778\pm 0.036$ & $0.356^{+0.018}_{-0.021}$ & $0.848^{+0.044}_{-0.050}$\\[0.5ex]
        3. $k_{\rm max}=0.2\,{\rm Mpc}^{-1}$ & $0.762\pm 0.032$ & $0.339\pm 0.017$ & $0.810\pm 0.041$\\[0.5ex]
        4. $G<20$ & $0.752\pm 0.043$ & $0.335\pm 0.020$ & $0.795^{+0.048}_{-0.054}$\\[0.5ex]
        5. ${\rm 1\,\,bin}$ & $0.745^{+0.028}_{-0.035}$ & $0.324^{+0.015}_{-0.019}$ & $0.774^{+0.035}_{-0.045}$\\[0.5ex]
        6. ${\rm Low}$-$z$ & $0.831\pm 0.046$ & $0.336^{+0.020}_{-0.022}$ & $0.879\pm 0.055$\\[0.5ex]
        7. ${\rm High}$-$z$ & $0.680\pm 0.051$ & $0.344^{+0.020}_{-0.024}$ & $0.729\pm 0.063$\\[0.5ex]
        8. $b(z)\,\,{\rm B2}$ & $0.743\pm 0.032$ & $0.337^{+0.016}_{-0.018}$ & $0.787\pm 0.040$\\[0.5ex]
        9. $b(z)\,\,{\rm B3}$ & $0.769\pm 0.034$ & $0.354\pm 0.019$ & $0.835\pm 0.043$\\[0.5ex]
        10. ${\rm Shot\,\,noise\,\,marg.}$ & $0.764\pm 0.041$ & $0.342\pm 0.019$ & $0.815\pm 0.045$\\[0.5ex]
        11. $p(z)\,\,{\rm stack}$ & $0.766\pm 0.034$ & $0.344\pm 0.017$ & $0.820\pm 0.042$\\[0.5ex]
        12. ${\rm Clust.}\,z,\,\,{\rm High}$-$z$ & $0.701\pm 0.051$ & $0.344^{+0.020}_{-0.025}$ & $0.751^{+0.058}_{-0.066}$\\[0.5ex]
        13. ${\rm No\,\,deproj}$ & $0.769\pm 0.035$ & $0.345\pm 0.018$ & $0.823\pm 0.045$\\[0.5ex]
        14. ${\rm Low}\,\,s$ & $0.795\pm 0.036$ & $0.337\pm 0.016$ & $0.843\pm 0.045$\\[0.5ex]
        15. $\kappa_{\rm pol}+{\rm High}$-$z$ & $0.86\pm 0.11$ & $0.348^{+0.023}_{-0.028}$ & $0.92\pm 0.13$\\[0.5ex]
        16. $\kappa_{\rm pol}$ & $0.814\pm 0.042$ & $0.338\pm 0.018$ & $0.864\pm 0.050$\\[0.5ex]
        17. $\theta_{\rm LSS}\,\,{\rm prior}$ & $0.756^{+0.030}_{-0.034}$ & $0.3163\pm 0.0079$ & $0.777\pm 0.033$\\[0.5ex]
        18. $H_0\,\,{\sl Planck}$ & $0.756\pm 0.033$ & $0.356^{+0.023}_{-0.028}$ & $0.823\pm 0.045$\\[0.5ex]
        19. $H_0\,\,{\rm SH0ES}$ & $0.792\pm 0.036$ & $0.316^{+0.021}_{-0.027}$ & $0.812\pm 0.046$\\[0.5ex]
        \hline\hline
      \end{tabular}
    \end{center}
    \caption{Constraints (marginalised mean and $68\%$ confidence level interval) on $(\sigma_8,\Omega_m,S_8)$ for the different cases studied here. Our fiducial result, described in Section \ref{ssec:res.fid}, is shown in the first row. All other rows show the results found in the various robustness tests described in Section \ref{ssec:res.rob}.}\label{tab:results}
  \end{table}
    \begin{figure}
      \centering
      \includegraphics[width=0.99\textwidth]{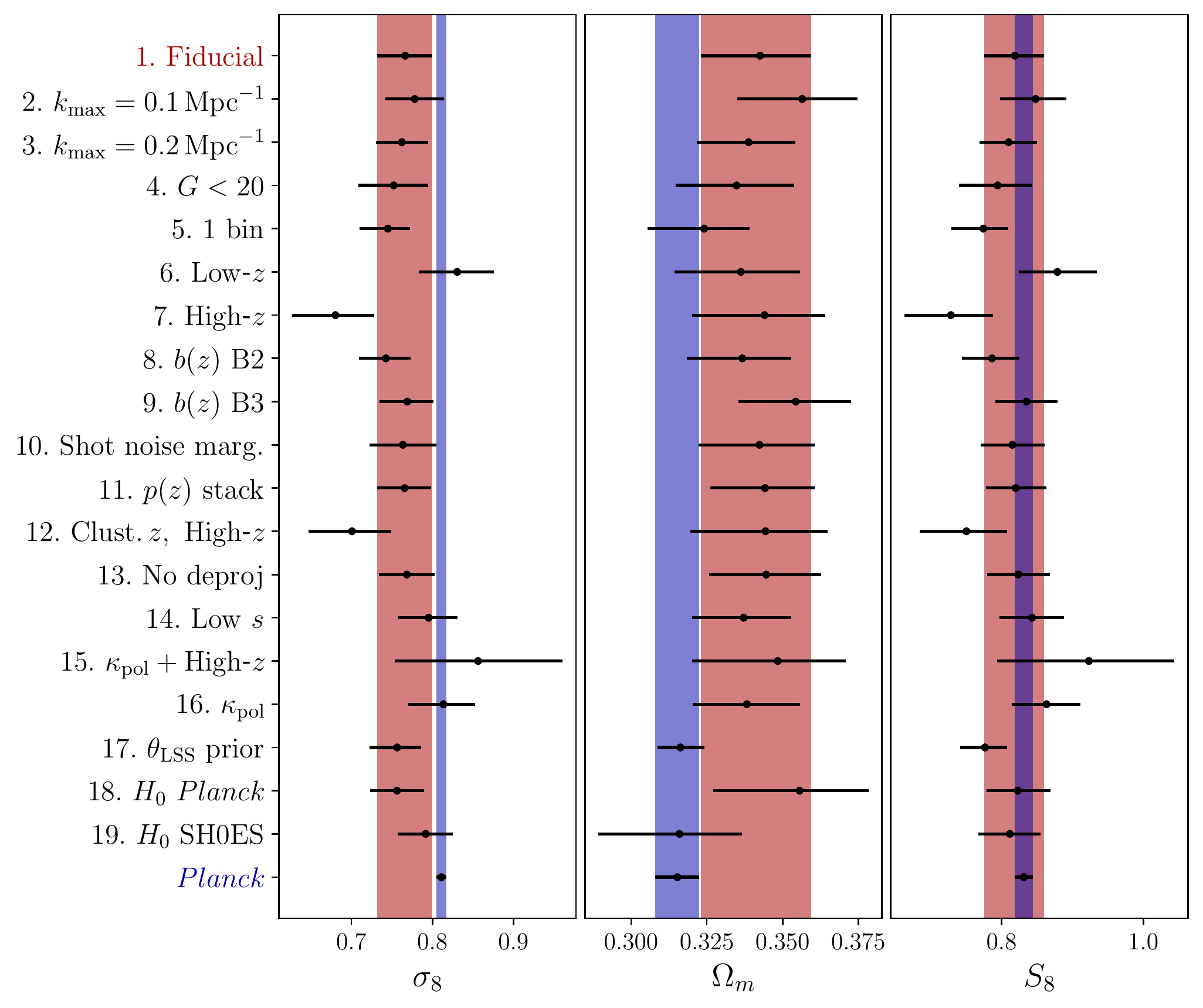}
      \caption{Visual representation of the results shown in Table \ref{tab:results}. Together with the results found in the different robustness described in Section \ref{ssec:res.rob}, we show the constraints found in our fiducial analysis (first row, and salmon band), and those found by \planck (last row and blue band).}\label{fig:results_tests}
    \end{figure}
    In this section we study the robustness of these results against the various analysis choices made, and potential sources of systematic uncertainty. Table \ref{tab:results} lists the final parameter constraints found for the different alternatives explored here, together with our fiducial measurements. The constraints on $(\sigma_8,\Omega_m,S_8)$ from these cases are also shown in Fig. \ref{fig:results_tests}. In all these cases we find a reasonably good fit to the data with $\chi^2$ PTE values ranging between 0.04 and 0.5.

    \subsubsection{Internal consistency}\label{sssec:res.rob.internal}
      \begin{figure}
        \centering
        \includegraphics[width=0.8\textwidth]{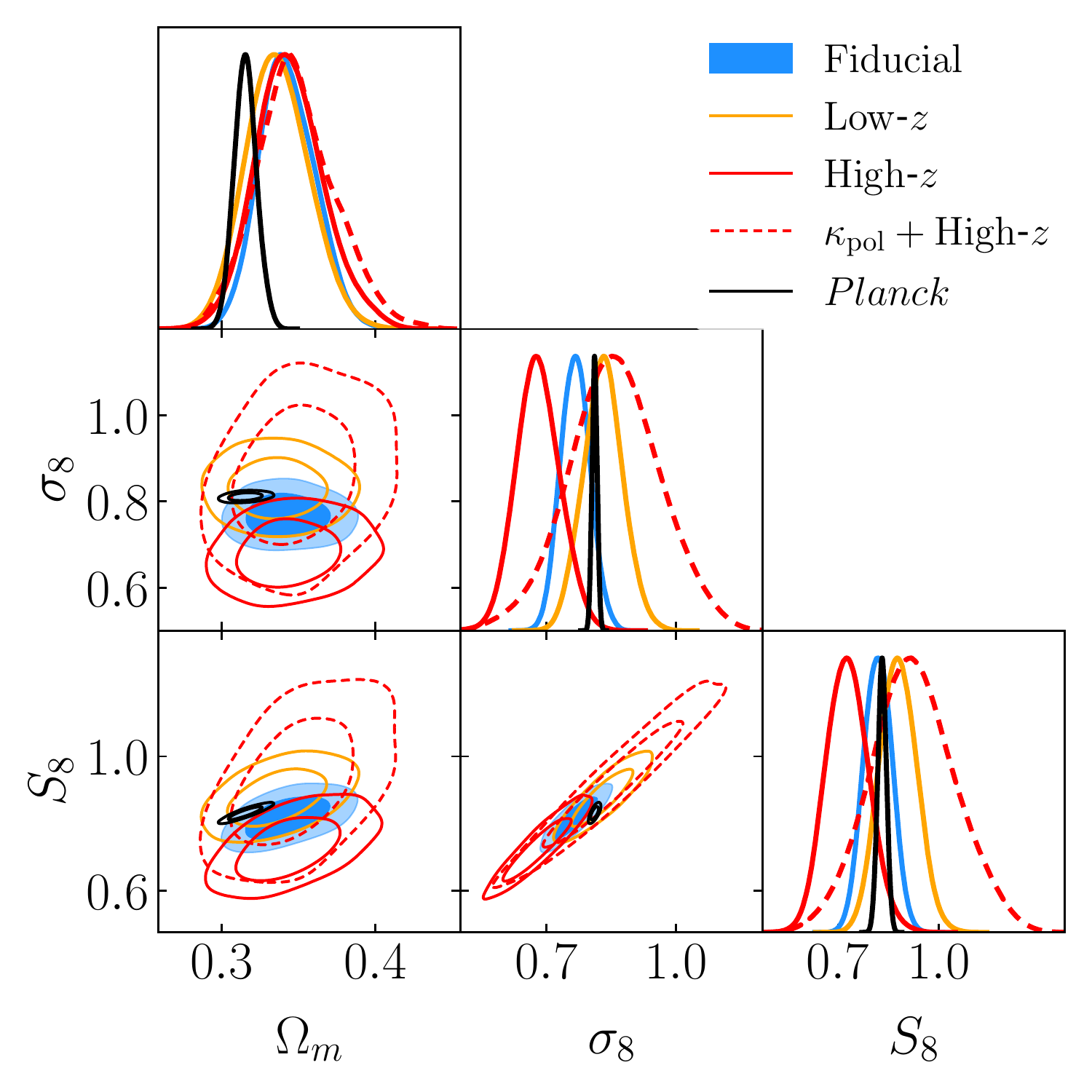}
        \caption{Constraints on $\Omega_m$, $\sigma_8$, and $S_8$ found in our fiducial analysis (blue contours), in comparison with those found using only each of the two \Quaia redshift bins independently (solid orange and red contours for the Low-$z$ and High-$z$ bins respectively). The \planck constraints are shown in black. The higher redshift bin recovers a value of $\sigma_8$ that is smaller than that found by \planck at the $2.5\sigma$ level, and is hence the main driver of the slightly lower value of $\sigma_8$ found in our fiducial analysis. The dashed red contours show the constraints found with the High-$z$ when cross-correlating with the polarization-only CMB lensing map (which is more reliable against extragalactic foreground contamination in the CMB map), as presented in Section \ref{sssec:res.rob.kappa_pol}. The $\sim$1.5$\sigma$ upwards shift in $\sigma_8$ observed is a hint that the low $\sigma_8$ found with the fiducial CMB $\kappa$ map may be caused by extragalactic foreground contamination in that map.}\label{fig:triangle_bins}
      \end{figure}
      We begin by exploring whether we obtain consistent results from different sub-sections of our data:
      \begin{itemize}
        \item We explore the impact of our {\bf small-scale cut} ($k_{\rm max}=0.15\,\iMpc$) on the final constraints by repeating our analysis imposing both more and less conservative cuts. Rows 2 and 3 in Table \ref{tab:results} and Fig. \ref{fig:results_tests} show the constraints found for $k_{\rm max}=0.1\,\iMpc$ and $k_{\rm max}=0.2\,\iMpc$. In both cases we recover constraints that are compatible with our fiducial measurements within $\sim0.5\sigma$, with a $\sim 6\%$ increase and reduction in the final uncertainties for the more and less conservative cuts respectively. Our results are thus robust to this choice. This also reinforces the validity of the simple linear bias model used in our analysis, since the presence of large non-linear corrections would lead to significant changes in the final constraints when adding or removing smaller scales.
        \item We repeat the analysis for a {\bf brighter quasar sample}, cutting at $G<20$. This brighter sample is potentially less sensitive to sky contamination and photo-$z$ mis-estimation than the fiducial catalog used in our analysis. For this sample, the magnification bias slope is given by Eq. \ref{eq:sG20}. As shown in row 4 of Table \ref{tab:results} and Fig. \ref{fig:results_tests}, the resulting parameter constraints are fully compatible with the fiducial case, with a $\sim30\%$ increase in the final uncertainties, caused by the higher shot noise of this sample.
        \item We obtain cosmological constraints from all \Quaia sources combined into a {\bf single redshift bin} (with the redshift distribution shown in black in Fig. \ref{fig:dndzs}). This test allows us to check for the impact of any potential mis-estimation of the intermediate-redshift tails of the redshift distributions in the 2-bin case, as well as increased sky contamination caused by the additional cut in redshift. The results, shown in row 5 of Table \ref{tab:results} and Fig. \ref{fig:results_tests}, are largely compatible with our fiducial constraints, with a shift of less than $1\sigma$ in both $\sigma_8$ and $\Omega_m$, and barely any change in the final uncertainties.
        \item Finally, we repeat our analysis for each of the {\bf two redshift bins separately}. The constraints are shown in rows 6 and 7 of Table \ref{tab:results} and Fig. \ref{fig:results_tests}, and the corresponding 2D contours are also shown separately in Fig. \ref{fig:triangle_bins}. The cosmological parameters inferred from each redshift bin are compatible with one another, although the High-$z$ bin favours a value of $\sigma_8$ that is $2.2\sigma$ lower than the Low-$z$ bin, $1.5\sigma$ lower than our fiducial constraints, and $2.5\sigma$ lower than \planck. We thus find that the slightly lower value of $\sigma_8$ found in our analysis in comparison with \planck is mostly driven by the higher redshift bin. We will therefore scrutinise the results obtained from this bin further in the following sections.
      \end{itemize}

    \subsubsection{Bias model}\label{sssec:res.rob.bz}
      \begin{figure}
        \centering
        \includegraphics[width=0.9\textwidth]{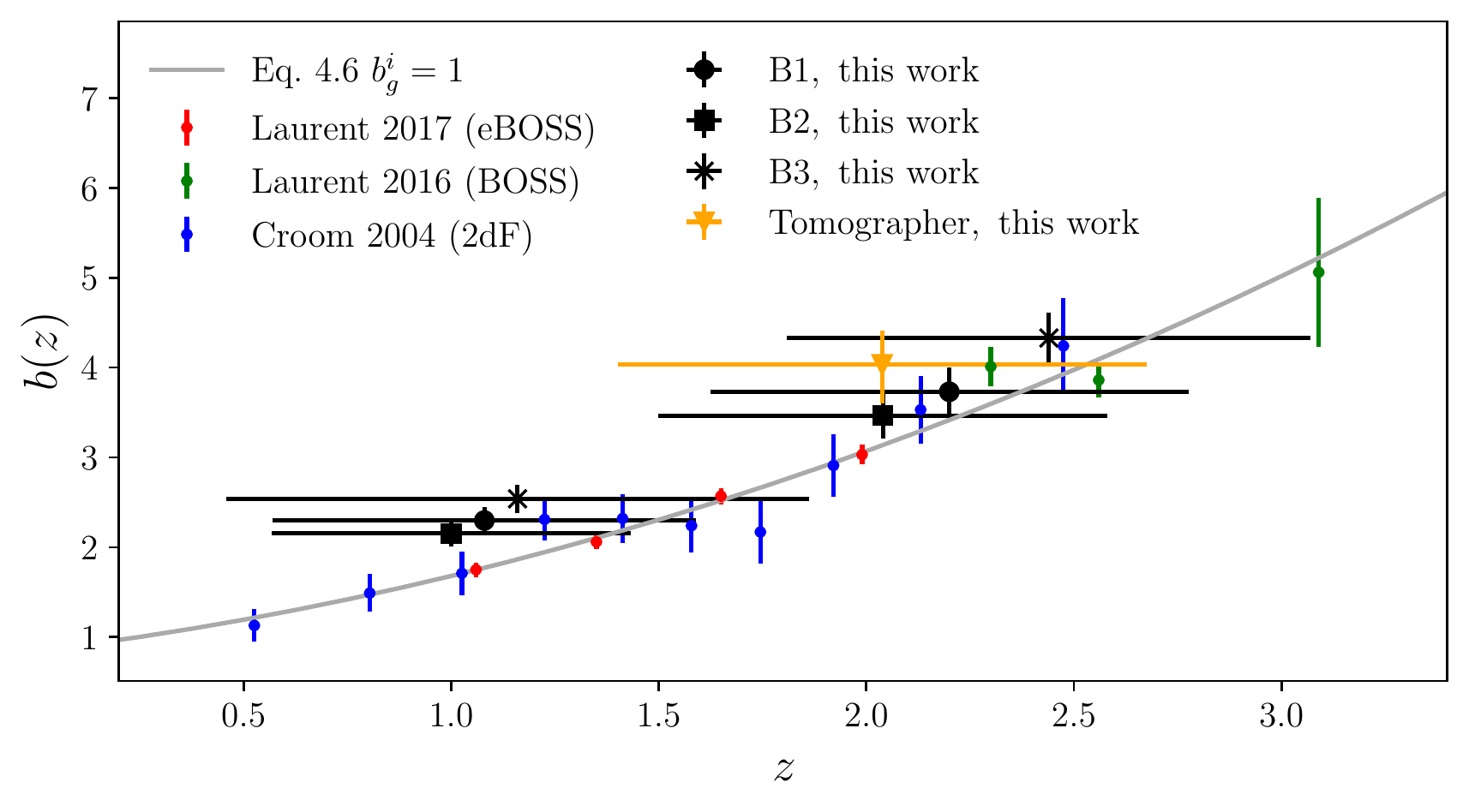}
        \caption{Measurements of the bias $b(z)$ for various quasar samples: the 2dF QSO Redshift Survey \cite{astro-ph/0409314} (blue), BOSS \cite{1602.09010} (green), and eBOSS \cite{1705.04718} (red). The grey line shows the fiducial bias evolution model used in our analysis, with the amplitude parameter $b_g$ fixed to 1, to reproduce the best-fit of \cite{1705.04718} for eBOSS. The bias values measured in this analysis for the \Quaia sample are shown in black for the three different bias evolution models considered (models B1, B2, and B3, described in Section \ref{sssec:res.rob.bz}). The value found when using {\rm Tomographer} to constrain the redshift dependence of the product $b(z)p(z)$ is shown in orange. Our constraints are in rough agreement with previous measurements. The systematically higher value of the quasar bias found for \Quaia is consistent with it being slightly brighter than the spectroscopic samples shown in the figure.}\label{fig:bevo}
      \end{figure}
      Our fiducial constraints assume the specific redshift dependence of the quasar bias in Eq. \ref{eq:bzfid}. For concreteness, we will label this {\bf Model B1} in this section. To quantify the dependence of our results on the assumed evolution of the quasar bias, we repeat our analysis for two other models:
      \begin{itemize}
        \item {\bf Model B2:} the model of \cite{1405.4315}, which is significantly steeper, and likely ruled out by current data:
        \begin{equation}
          b(z)=b_g^i\,\left[1+\left(\frac{1+z}{2.5}\right)^5\right],
        \end{equation}
        \item {\bf Model B3:} a constant bias model ($b(z)=b_g^i$), also ruled out by current data if applied at all redshifts.
      \end{itemize}
      The constraints found with these two alternative models are shown in rows 8 and 9 of Table \ref{tab:results} and Fig. \ref{fig:results_tests}. We observe that models with a steeper bias evolution lead to lower values of both $\sigma_8$, and $\Omega_m$. Nevertheless, for both models, B2 and B3, which encompass the degree of evolution exhibited by existing measurements, the parameter shifts are mild (less than $1\sigma$). Hence, our results do not depend strongly on this choice.

      It is useful to place our constraints on quasar bias in the context of previous measurements. For the three bias evolution models studied, we calculate the effective bias in each of the two redshift bins as
      \begin{equation}
        b_{\rm eff}=\frac{\int dz\,p(z)\,b(z)}{\int dz\,p(z)},
      \end{equation}
      and assign to it an effective redshift given by the mean $z$ in each bin weighted by the product $b(z)p(z)$. The results are shown in Fig. \ref{fig:bevo} for all three models (black symbols), together with the measurements of \cite{astro-ph/0409314,1602.09010,1705.04718} (blue, green, and red points), and our fiducial model (Eq. \ref{eq:bzfid}, with $b_g^i=1$). Reassuringly, our measurements are in rough agreement with other quasar samples. In particular, we find no strong evidence of the lower quasar bias at high redshifts ($z\gtrsim2$) reported by previous analyses using CMB lensing cross-correlations \cite{1706.04583,1712.02738,2011.01234,2305.07650}. As discussed in Section \ref{sssec:meth.gc.magbias}, the impact of magnification bias, an effect that the CMB lensing cross-correlation is particularly sensitive to, is likely negligible in our sample. Finally, it is worth noting that, although our bias constraints lie somewhat above most previous measurements in both redshift bins, this is not entirely surprising, as our sample is different (brighter, in fact) from the spectroscopic samples targeted by previous measurements.

      Our fiducial bias model is purely linear, and ignores contributions from higher-order terms. This assumption is likely valid given the relatively conservative scale cuts used \cite{2105.13546,2111.10966,2304.00705}, and we are further reassured by the robustness of our constraints to changes in these scale cuts (as shown in Section \ref{sssec:res.rob.internal}). However, we have assumed that the stochastic term in Eq. \ref{eq:blinmod} is completely described by Poisson statistics, and can be subtracted analytically. Since a stochastic term would affect the quasar auto-correlation on all harmonic scales, we test this assumption by repeating our analysis varying the amplitude of the shot-noise contribution to $C_\ell^{gg}$ with a $10\%$ prior. This also allows us to test for any misestimation of the purely-Poisson noise level. The result is shown in row 10 of Table \ref{tab:results} and Fig. \ref{fig:results_tests}. The mean value of all parameters remains virtually unchanged, the only effect being a $\sim10-20\%$ increase in the final uncertainties. The data thus shows no preference for additional flat contribution to $C_\ell^{gg}$, and its inclusion causes no shift on the final parameter constraints.

    \subsubsection{Redshift uncertainties}\label{sssec:res.rob.pz}
      \begin{figure}
        \centering
        \includegraphics[width=0.9\textwidth]{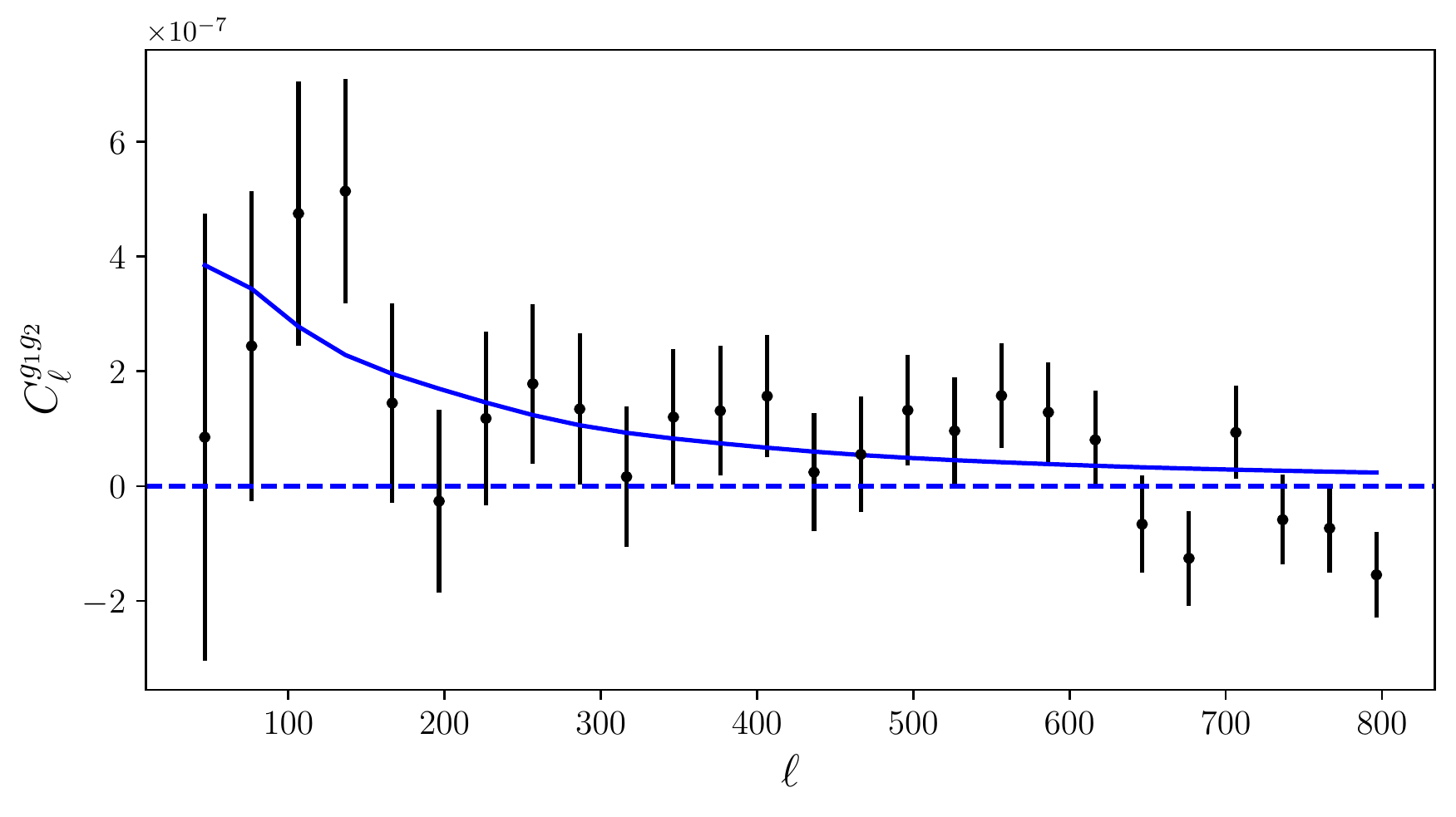}
        \caption{Cross-correlation between the two \Quaia redshift bins (black points with error bars), and prediction from our best-fit model. This cross-correlation is sensitive to the overlap between the redshift distributions of both bins, shown in Fig. \ref{fig:dndzs}. Our best-fit model provides a good fit to this cross-correlation, which was not used in that analysis.}\label{fig:clg1g2}
      \end{figure}
      To test the robustness of our constraints to a mis-calibration of the quasar redshift distributions, we repeat our analysis assuming the $p(z)$ calculated via PDF stacking, as described in Section \ref{sssec:meth.gc.pz}. The results are shown in row 11 of Table \ref{tab:results} and Fig. \ref{fig:results_tests}. The constraints are virtually unchanged, which shows that our results are insensitive to mild changes in the redshift distributions (such as the height of the bump\footnote{The origin of this bump is discussed in more detail in \cite{quaia}, and is in fact robust to the choice of spectroscopic training sample.} at $z\sim2.25$ shown in Fig. \ref{fig:dndzs}).

      Another important systematic related to the $p(z)$ is a potential mis-estimate of the inter-bin tails of both samples. Driven by photometric redshift outliers, they control the amplitude of the redshift distribution, and hence the amplitude of the quasar auto-correlation. Fortunately, the cross-correlation between both bins is particularly sensitive to the details of the redshift overlap between them \cite{1912.08209,2010.00466}. This cross-correlation is shown in Fig. \ref{fig:clg1g2}, together with the theoretical prediction for our fiducial best-fit model. On scales $30\leq\ell\leq800$, the cross-correlation is detected at the $3.2\sigma$ level, and our best-fit model provides a good fit to it (${\rm PTE}=0.51$), without any additional parameters. This shows that the combination of best-fit bias parameters found in both bins, and the intermediate-redshift tails of their redshift distributions, are compatible with a complementary piece of the data vector that was not used in the likelihood.

      As shown in Section \ref{sssec:res.rob.internal}, the lower value of $\sigma_8$ we recover compared to \planck is largely driven by the auto- and cross-correlation of the High-$z$ quasar redshift bin. Since both the shape of the redshift distribution, and the evolution of the linear bias may depart from the fiducial models used here, particularly at high redshifts. (e.g. due to photo-$z$ outliers or a steeper-than-expected $b(z)$ relation), we perform one further check to test the robustness of the low $\sigma_8$ value measured for that sample. We made use of the clustering redshifts technique, as implemented in \tomog\footnote{\url{http://tomographer.org}}, to measure the product $b(z)p(z)$ up to an unknown normalization factor. \tomog uses a spectroscopic reference sample from SDSS comprising 2 million sources at redshifts $z\lesssim3$. Through its online interface, the code transforms an input catalog into a \hpix map at resolution $N_{\rm side}=2048$, imposes a high-pass filter removing scales larger than $\sim2^\circ$ to mitigate the impact of sky systematics, and calculates the correlations between the map and the reference sample in thin redshift bins on scales $r\in[2,8]\,{\rm Mpc}$. In combination with the reference sample auto-correlations, this is then used to provide constraints (mean and standard deviation) of $b(z)p(z)$ in each spectroscopic redshift bin.
      \begin{figure}
        \centering
        \includegraphics[width=0.8\textwidth]{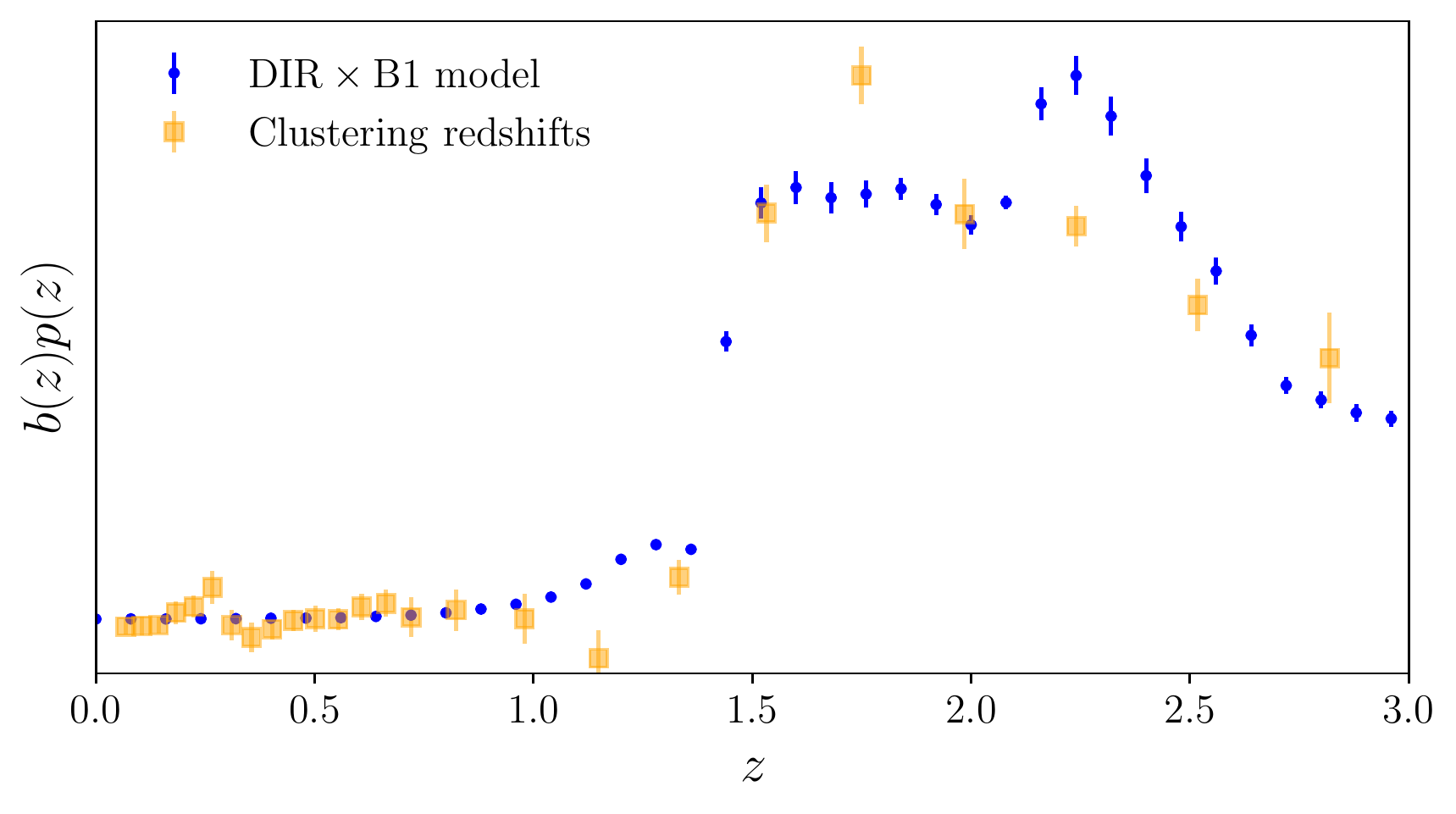}
        \caption{Product of the redshift distribution and linear bias of the High-$z$ \Quaia redshift bin. The blue points show the DIR estimate of the redshift distribution multiplied by our fiducial bias evolution model (Eq. \ref{eq:bzfid}). The orange points show the measurement of $b(z)p(z)$ made using the clustering redshifts approach of \tomog.}\label{fig:bpz}
      \end{figure}
      
      We use \tomog to estimate $b(z)p(z)$ for the High-$z$ bin in our sample. The redshift coverage of the reference spectroscopic sample in \tomog prevents a reliable determination of $b(z)p(z)$ beyond $z\sim3$, and thus, for this test, we impose an additional cut in the catalog at $\zquaia<3$. As with the DIR redshift distribution, we marginalise over the uncertainties in $b(z)p(z)$ obtained by \tomog using the analytical marginalisation method of \cite{2007.14989}. The cosmological constraints obtained from this test are shown in line 12 of Table \ref{tab:results} and Fig. \ref{fig:results_tests}. Compared with the constraints found with the High-$z$ bin using the DIR redshift distribution and our fiducial  bias evolution template, we observe only a small upwards shift in $\sigma_8$ ($\delta\sigma_8\simeq0.02$, less than $0.5\sigma$). Hence, the data still favours a lower value of $\sigma_8$ compared to \planck, although at a lower significance of $2.1\sigma$. This slightly better agreement with \planck could be simply a statistical fluctuation, it could be caused by the slightly lower effective redshift of the sample, or could be evidence of a departure from the quasar bias evolution model assumed here. The effective bias obtained from our MCMC chains for this bin using clustering redshifts is shown in orange in Fig. \ref{fig:bevo}, and agrees well with the values found for the three other bias evolution models explored. Fig. \ref{fig:bpz} shows the values of $b(z)p(z)$ recovered from \tomog in orange, together with the DIR-calibrated $p(z)$ multiplied by the B1 bias evolution model (Eq. \ref{eq:bzfid}). Both predictions for the product $b(z)p(z)$ are qualitatively similar in terms of the redshift support and high-redshift tails. The small difference observed in the final constraints might be caused by the enhancement in the bump at $z\sim2.25$ in the DIR distribution caused by the strong bias evolution at that redshift. Overall, however, we find that both predictions yield highly compatible constraints for the High-$z$ bin, and hence, when combined with the results presented in this section and Section \ref{sssec:res.rob.bz}, it is unlikely for the lower value of $\sigma_8$ found in this bin to be caused by mis-modelling of the sample's redshift distribution or its bias evolution.

    \subsubsection{Sky contamination}\label{sssec:res.rob.sky}
      \begin{figure}
        \centering
        \includegraphics[width=0.75\textwidth]{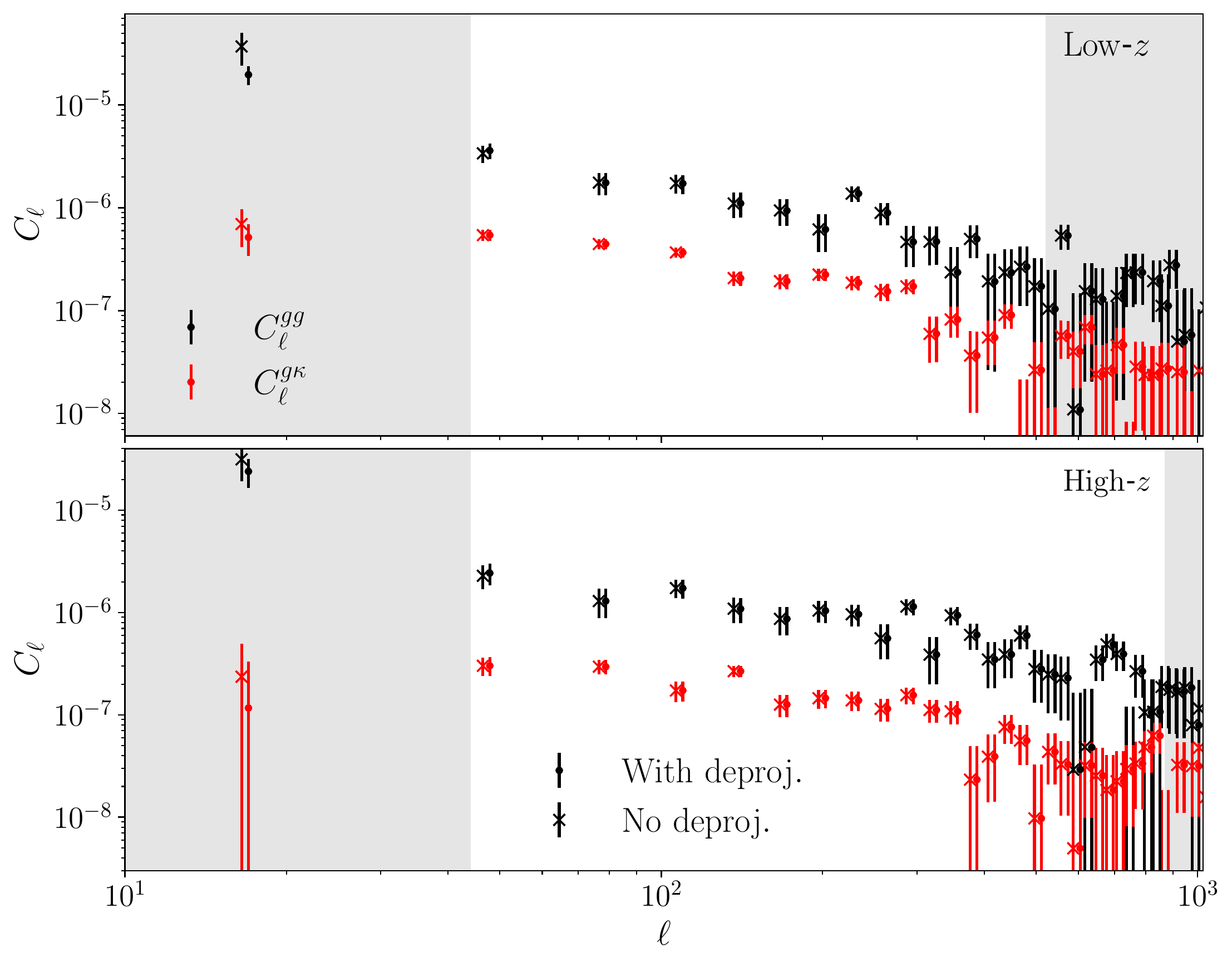}
        \includegraphics[width=0.75\textwidth]{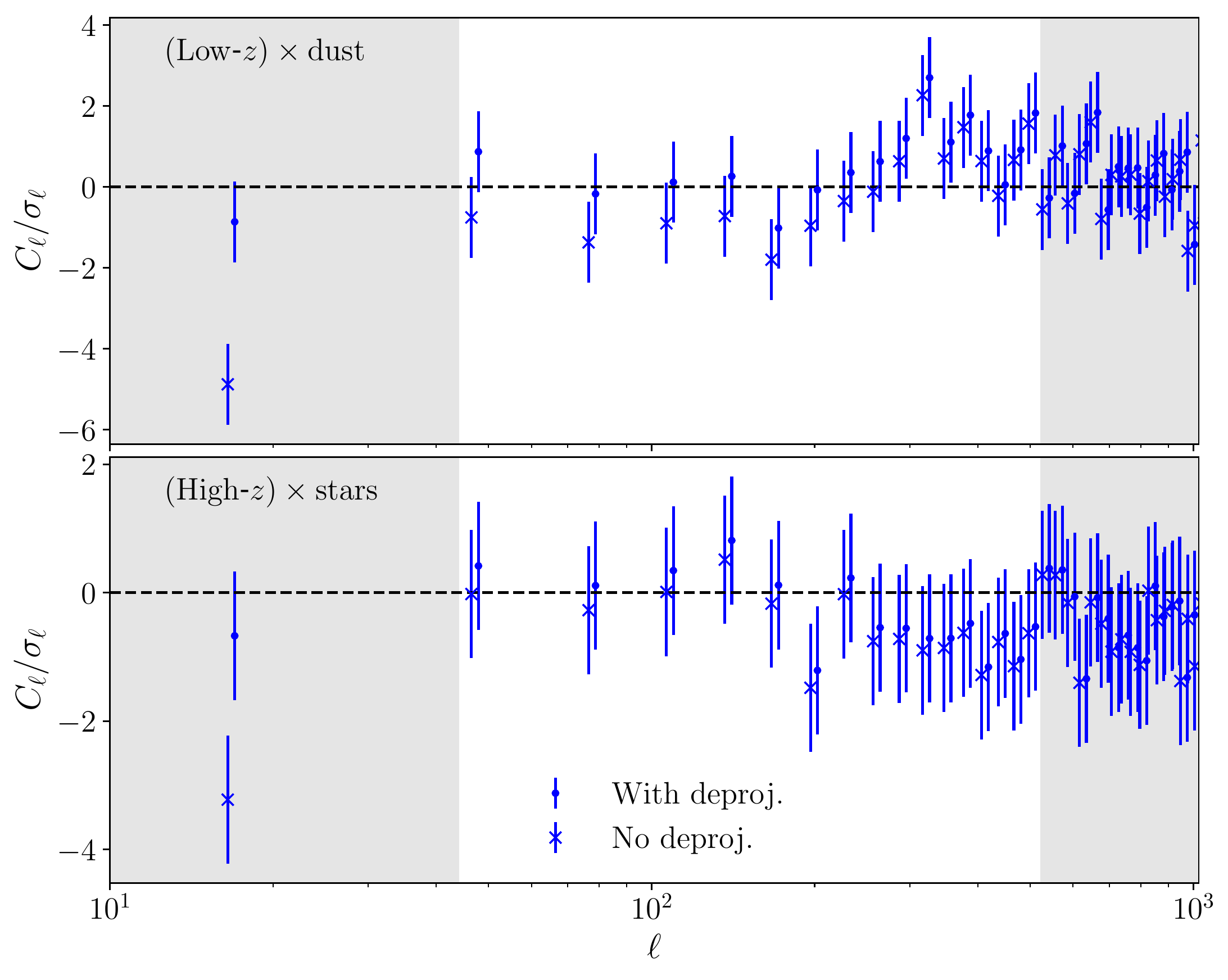}
        \caption{{\sl Top:} quasar-quasar (black) and quasar-$\kappa$ (red) angular power spectra for the two redshift bins used in this analysis (top and bottom subpanels). The measurements made using linear deprojection to remove residual sky contamination are shown as points with error bars, while the crosses show the results found without deprojection. The impact of residual contamination is mostly relevant for the first bandpower of the auto-correlation, which we discard in our analysis. {\sl Bottom:} example cross-correlations between the quasar overdensity maps and the most relevant sky contaminants for each of the \Quaia redshift bins. Results are shown with and without linear deprojection (points and crosses respectively), divided by their statistical uncertainties to better highlight significant departures from zero correlation. Evidence of contamination is mostly present in the first bandpower, in agreement with the results of the top panel.}\label{fig:cls_syst}
      \end{figure}
      The presence of residual sky contaminants in the quasar overdensity maps could cause a bias in the final cosmological constraints. For instance, if purely additive, contamination would raise the amplitude of the quasar auto-correlation, leaving the cross-correlation intact. When fitting for bias and $\sigma_8$, this would then lead to an upwards shift in the quasar bias (to accommodate the higher auto-correlation) and, correspondingly, a downwards shift in $\sigma_8$ (to compensate for the larger bias in the cross-correlation). To quantify the robustness of our results to the presence of residual contamination, we repeat the analysis skipping the final linear deprojection step when computing all power spectra. As a reminder, this step was meant to ensure that any small residual contamination from the known sky systematics was removed from the maps. The top panel of Fig. \ref{fig:cls_syst} shows the four angular power spectra used in our analysis with (points) and without (crosses) performing the final linear deprojection step. We see that linear deprojection has a significant effect on the first bandpower, which is discarded in our analysis, and less of an effect on smaller scales. The constraints found without linear deprojection are shown in row 13 of Table \ref{tab:results} and Fig. \ref{fig:results_tests}. We observe only negligible shifts in the best-fit parameters ($\lesssim0.1\sigma$), and thus conclude that our results are likely robust against residual contamination from the known sky systematics.

      To further understand this result, we compute the cross-correlation between the quasar overdensity maps and the 4 contaminant maps described in Section \ref{sssec:meth.gc.map}. The detection of a non-zero correlation would constitute evidence of residual contamination in the maps. We find that, without linear deprojection, the quasar overdensity maps exhibit large-scale residual contamination, particularly from dust extinction and stars. This contamination disappears after deprojection (unsurprisingly, since linear deprojection is precisely intended to cancel these correlations at the pixel level). An example of this is shown in the lower panel of Fig. \ref{fig:cls_syst} for the two \Quaia bin--systematics combinations exhibiting the most contamination. Although linear deprojection modifies some of the cross-correlations on all scales, they are only significantly different from zero in the first bandpower without deprojection, in agreement with the results found above.

   \subsubsection{Magnification bias}\label{sssec:res.rob.magbias}
      As discussed in Section \ref{sssec:meth.gc.magbias}, the magnification bias slope of our fiducial quasar sample, measured with two different methods, is close to $s\simeq0.4$, and hence our measurements are largely unaffected by magnification bias. To quantify the sensitivity of our final constraints to a potential mis-estimation of this quantity, we repeat our analysis fixing the number counts slope to the value found for the DESI quasar sample, $s=0.2764$ \cite{2305.07650}. This value is in rough agreement with previous estimates for similar spectroscopic quasar samples (e.g. in BOSS and eBOSS). The result from this exercise, shown in row 14 of Table \ref{tab:results} and Fig. \ref{fig:results_tests}, leads to a $<1\sigma$ upwards shift in $\sigma_8$, and almost no change in the $\Omega_m$ constraint. It is worth noting that this value of $s$ is $\gtrsim3\sigma$ away from our estimate for \Quaia (see Section \ref{sssec:meth.gc.magbias}), and is therefore not supported by the data. Nevertheless, the impact on our constraints of such a significant misestimate of $s$ is small.
      \begin{figure}
        \centering
        \includegraphics[width=0.47\textwidth]{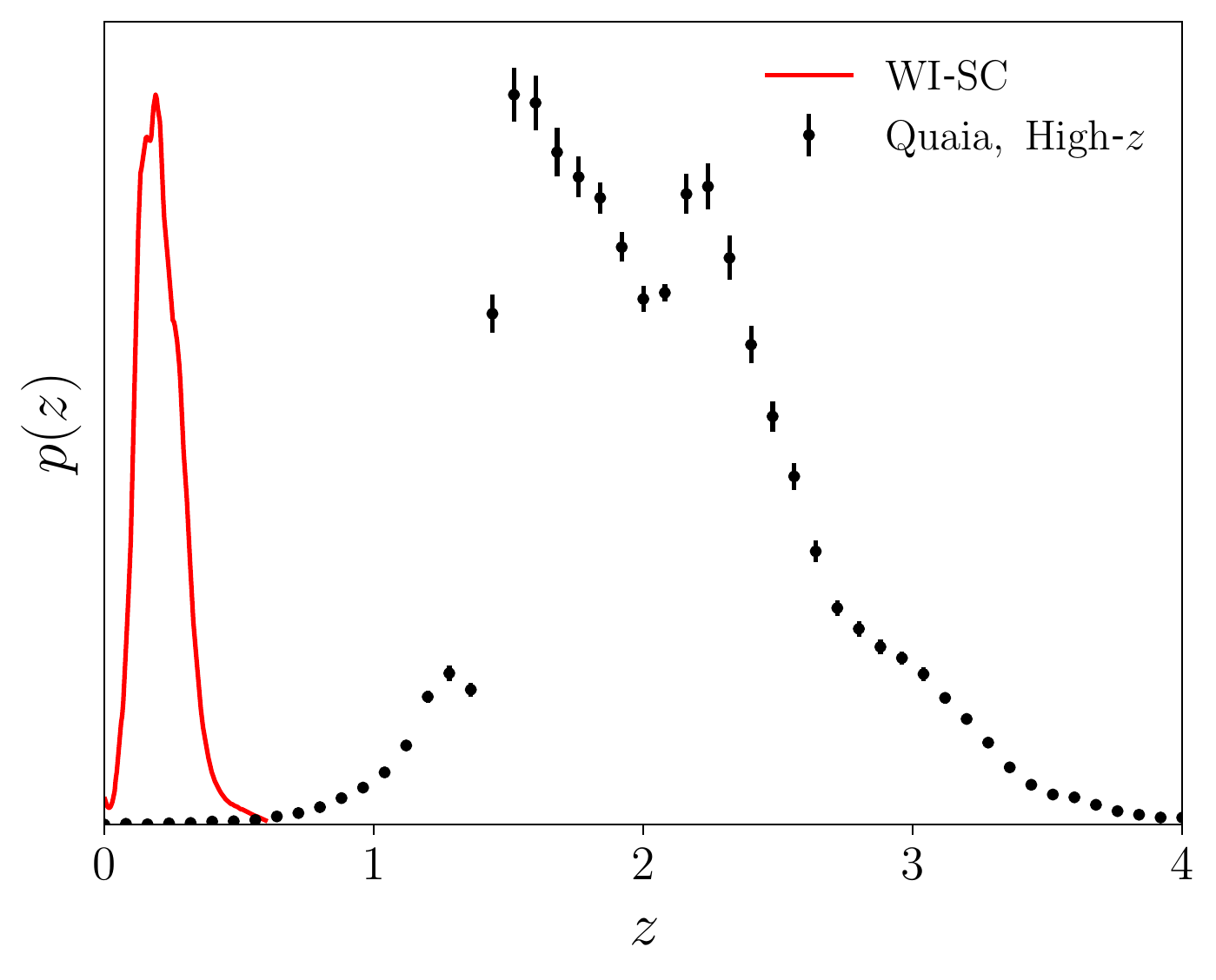}
        \includegraphics[width=0.51\textwidth]{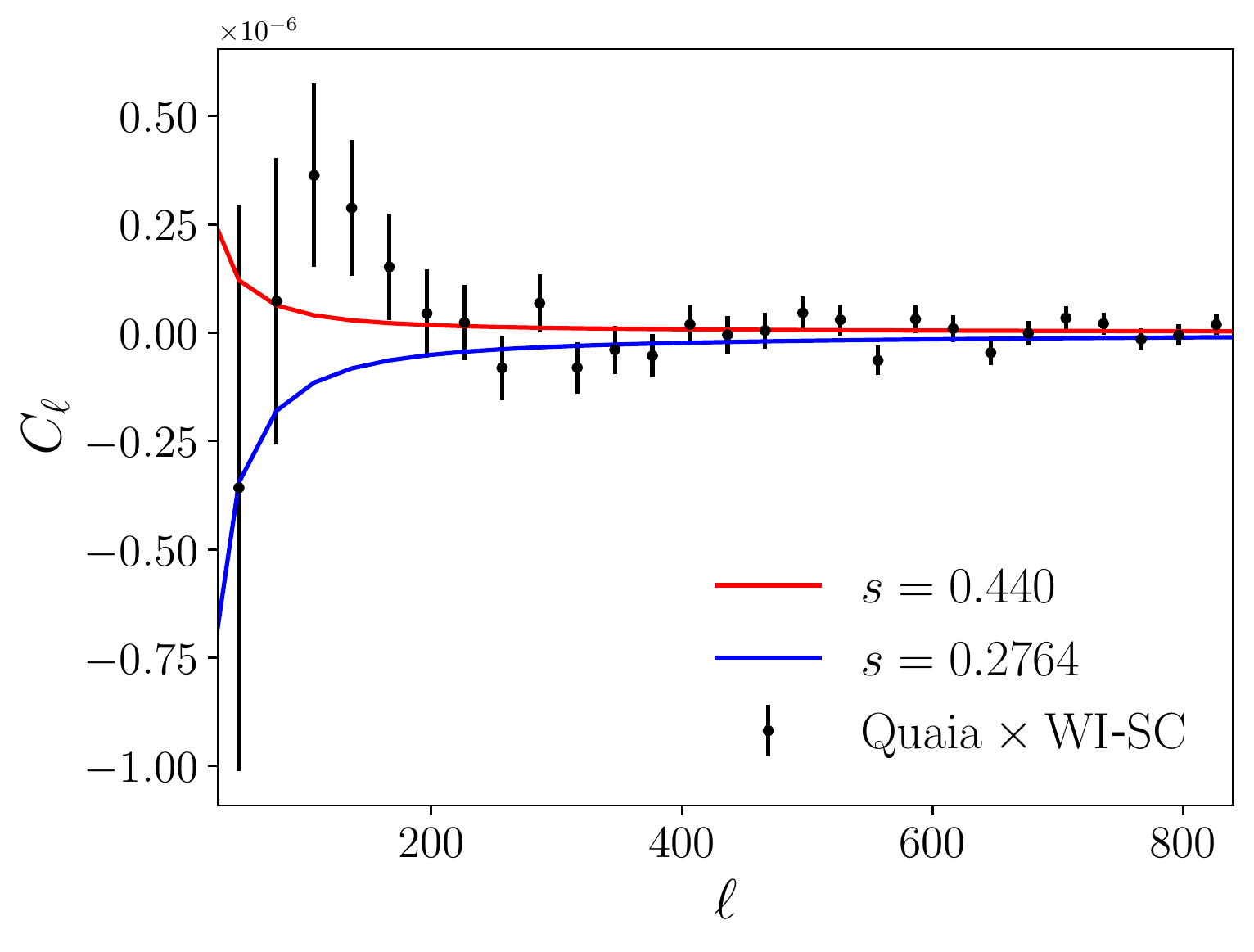}
        \caption{{\sl Left:} redshift distribution of the High-$z$ \Quaia bin (points with error bars), and of the WISE-SuperCOSMOS sample we cross-correlate it with to validate our estimate of the magnification bias slope. {\sl Right:} cross-correlation between the High-$z$ \Quaia bin and the low-redshift WISE-SuperCOSMOS sample. The solid lines show the predictions for the best-fit value of the magnification bias slope $s$ (red), and for the value found by \cite{2305.07650} in the DESI quasar sample (blue). The cross-correlation is able to rule out a value of $s$ as low as that found in the DESI sample at $\sim3.5\sigma$.}\label{fig:wisc}
      \end{figure}

      As an additional test to validate our estimate of $s$ for the High-$z$ bin (which shows the largest tension with \planck), we measure the cross-correlation between this bin and a low-redshift sample of galaxies selected from the WISE-SuperCOSMOS catalog (WI-SC, \cite{1607.01182}). In particular, we select 14.7 million galaxies with photometric redshifts $z_p\in[0.1,0.3]$. The methods used to analyse this sample are described in detail in \cite{1909.09102}, and we will not repeat them here for brevity. As shown in the left panel of Fig. \ref{fig:wisc}, both samples have negligible overlap in redshift, and thus, any detection of the cross-correlation between them would be largely driven by magnfication bias with $s\neq0.4$. 
     
      Fixing cosmological parameters to the best-fit values found by \planck \cite{1807.06209}, we determine the linear bias of the WI-SC sample from its auto-correlation on large scales to be $b_g=1.1$, in good agreement with \cite{1805.11525}. Fixing the galaxy bias to this value, and the quasar bias to the value found for the High-$z$ bin in our fiducial analysis, we use the cross-correlation between both samples to constrain the value of $s$. The right panel of Fig. \ref{fig:wisc} shows this cross-correlation, together with the prediction for the best-fit value of the number counts slope (see Eq. \ref{eq:sclust}) in red, and that for the value of $s$ found by \cite{2305.07650} for DESI in blue. The value of $s$ obtained from this cross-correlation is
      \begin{equation}\label{eq:sclust}
        s(z>1.47)=0.440\pm0.047,
      \end{equation}
      and provides a good fit to the data ($\chi^2=26.7$ for 26 degrees of freedom). This value is in good agreement with that found with two different methods in Section \ref{sssec:meth.gc.magbias}, whereas the DESI value lies $\sim3.5\sigma$ away from it. It is thus unlikely for magnification bias to be the cause of the low $\sigma_8$ value found from the higher-redshift quasar sample.

    \subsubsection{Foreground contamination in CMB lensing}\label{sssec:res.rob.kappa_pol}
      \begin{figure}
        \centering
        \includegraphics[width=0.9\textwidth]{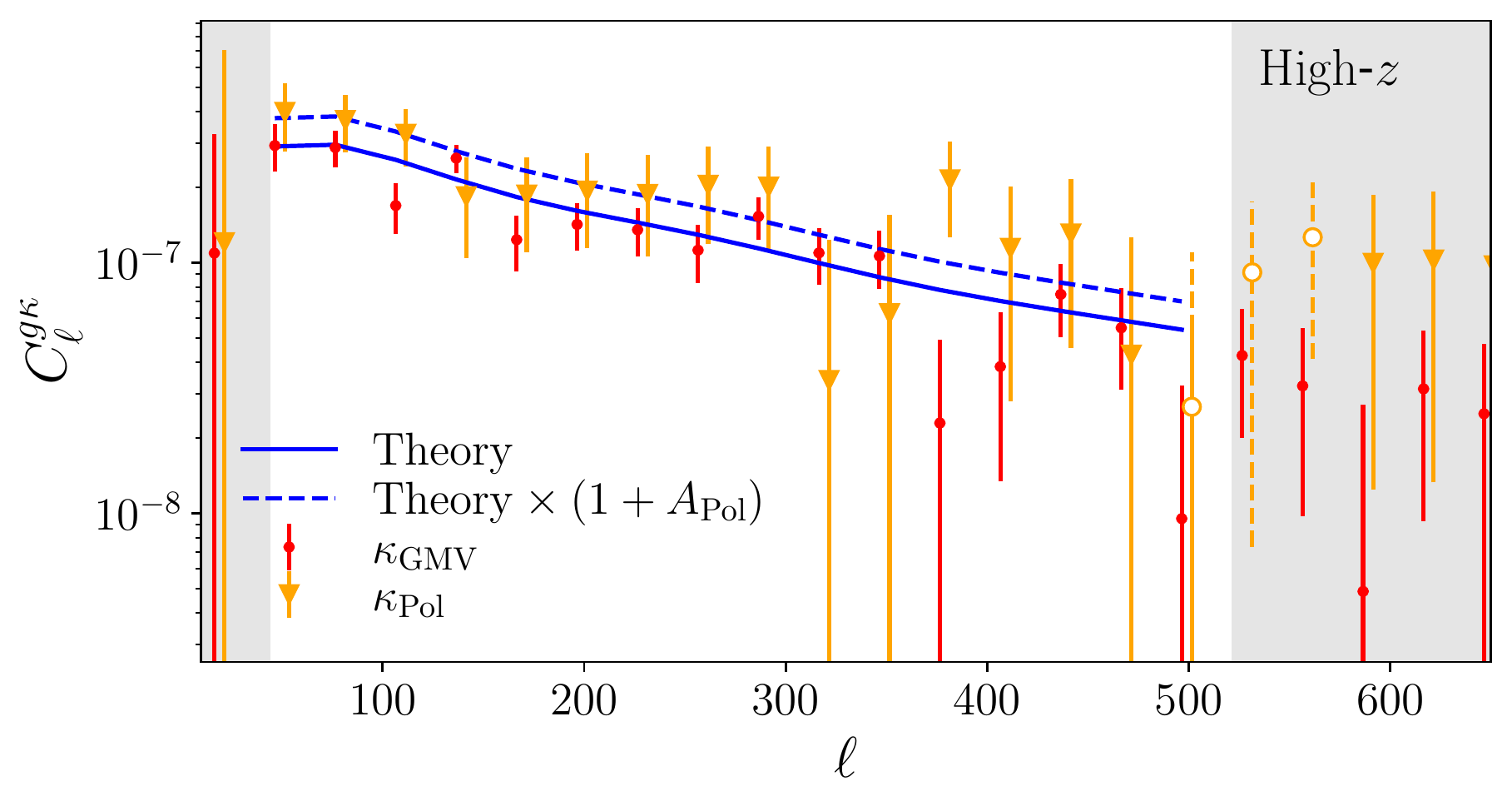}
        \caption{Cross-correlation between the High-$z$ \Quaia redshift bin, and maps of the CMB convergence. Results are shown for the fiducial generalised minimum variance lensing map (red points), and for the polarization-only map (orange triangles). When the power spectrum is negative, we show its absolute value as empty circles with dashed error bars. The polarization-only data is systematically above the GMV estimator on scales $\ell\lesssim300$, although with larger uncertainties, with all deviations at the level of $\sim$1$\sigma$. The blue solid line shows the best-fit theoretical prediction found in our fiducial analysis. In turn, the dashed blue line shows the theoretical prediction resulting from multiplying the fiducial best-fit curve by a free amplitude and fitting it to the difference between both power spectra. The best-fit curve found this way is $\sim$30\% higher than our fiducial best fit, which is significant at the $\sim$2$\sigma$ level. This translates into a $\sim$1.5$\sigma$ upwards shift in $\sigma_8$ for this bin when using the polarization-only map, making it fully compatible with the \planck value.}\label{fig:kpol}
      \end{figure}
      So far we have mostly explored the impact of various systematics affecting the \Quaia sample. One of the most important sources of systematic uncertainty for cross-correlation studies, however, is contamination in the CMB lensing map, particularly from extragalactic foregrounds, tracing the same large-scale structure as the probe being cross-correlated.

      To quantify the robustness of our results against this potential systematic, we computed the cross-correlation of the \Quaia sample with an alternative map of the CMB convergence constructed using only polarization information (and therefore largely devoid of any extragalactic foreground contamination that is mainly unpolarized) and a PR3-like pipeline run on NPIPE maps. We note that lensing reconstruction based on polarization can be more prone to biases generated by Galactic foregrounds, such as Galactic dust emission, which might in turn correlate with uncorrected dust extinction in the \Quaia sample. However, \cite{beck2020} showed that dust biases are mitigated to a negligible level if lensing reconstruction is performed on a CMB map cleaned of foregrounds through component separation (such as those used for PR3 and PR4 lensing analyses), even for data much deeper than \planck.  The systematic deprojection performed on the \Quaia sample would reduce this effect further. Using the methods described below we find no indication of contamination in the cross-correlation with the Low-$z$ bin, and therefore we will limit our discussion to the High-$z$ bin, which displays the largest deviation in $\sigma_8$ with respect to \planck (see Section \ref{sssec:res.rob.internal}).

      Figure \ref{fig:kpol} shows the cross-correlation of the High-$z$ bin with our fiducial generalised minimum variance (GMV) $\kappa$ map in red, and with the polarization-only map $\kappa_{\rm Pol}$ in orange. Although, in each bandpower, both power spectra are compatible within $1\sigma$, we find that the polarization-only coss-correlation is systematically $\sim30\%$ higher on large scales ($\ell\lesssim300$). At the \planck sensitivity, the polarization-only reconstructed $\kappa$ map is significantly noisier than the GMV estimator (the errors in the cross-correlation are between $\sim$2 and $\sim$4 times larger on the scales explored here), and hence we find only mild evidence that this shift is indeed caused by contamination in the $\kappa$ map. To quantify this in a model-independent way, we first calculate the $\chi^2$ of the difference between both power spectra with respect to a null model\footnote{Note that the covariance of the difference between two variables $\Delta{\bf x}_{12}\equiv {\bf x}_1-{\bf x}_2$ is ${\sf C}_{11}+{\sf C}_{22}-2{\sf C}_{12}$, where ${\sf C}_{ij}\equiv{\rm Cov}({\bf x}_i,{\bf x}_j).$}. The PTE of the resulting $\chi^2$ value is $0.37$, and hence, by this metric, both power spectra are compatible with each other within $1\sigma$. To quantify the evidence for a coherent deviation between both spectra, we instead take the theoretical prediction for the best-fit parameters found in our fiducial analysis for this cross-correlations, and multiply it by a free amplitude $A_{\rm Pol}$. We fit this 1-parameter model to the difference between both power spectra on scales $30\leq\ell\leq600$, finding:
     \begin{equation}
       1+A_{\rm pol}=1.296\pm0.125.
     \end{equation}
     We thus find weak evidence, at the $2.3\sigma$ level, that the power spectrum computed with the polarization-only $\kappa$ map is $\sim30\%$ higher than that for the minimum-variance map. A more in-depth investigation of this issue, making use of the \planck CMB lensing simulations in the PR3 release, is described in Appendix \ref{app:kappa_pol}.

     Finally, we propagate the impact of this potential systematics to our final constraints. For the High-$z$ \Quaia bin, using the polarization-only $\kappa$ map, we find the cosmological constraints shown in row 15 of Table \ref{tab:results} and Fig. \ref{fig:results_tests}. These constraints are also shown as dashed blue lines in Fig. \ref{fig:triangle_bins}. We observe a $\sim$1.4$\sigma$ upwards shift in $\sigma_8$, which is now fully compatible with \planck within 1$\sigma$. Although the evidence is relatively weak (at the $\sim$2$\sigma$ level), this provides an attractive explanation for the low amplitude of our fiducial cross-correlation between CMB lensing and the High-$z$ bin with respect to the \planck prediction. The most likely source of contamination behind this effect would be the Cosmic Infrared Background (CIB), given its relevance at redshifts $\gtrsim2$, and the fact that the resulting bias to the CMB lensing map and power spectrum is expected to be negative (see Fig. 23 of \cite{1807.06210}). The cosmological constraints found using using the GMV $\kappa$ map with the Low-$z$ bin, and the polarization-only map with the High-$z$ bin, are shown in row 16 of Table \ref{tab:results} and Fig. \ref{fig:results_tests}. The main effect is an upwards shift of $\sigma_8$ by $\sim$1$\sigma$, and a $\sim$30\% increase in the final uncertainties caused by the significantly higher noise of the polarization-only map.

   \subsubsection{Impact of priors}\label{sssec:res.rob.prior}
     As explained in Section \ref{ssec:meth.like}, our data is not sufficient to break the degeneracy between $\Omega_m$ and $h$, and we have relied on low-redshift BAO data to pin down $\Omega_m$. To explore the dependence of our results on this choice, we have repeated our analysis with three alternative prior choices.

     \begin{figure}
       \centering
       \includegraphics[width=0.7\textwidth]{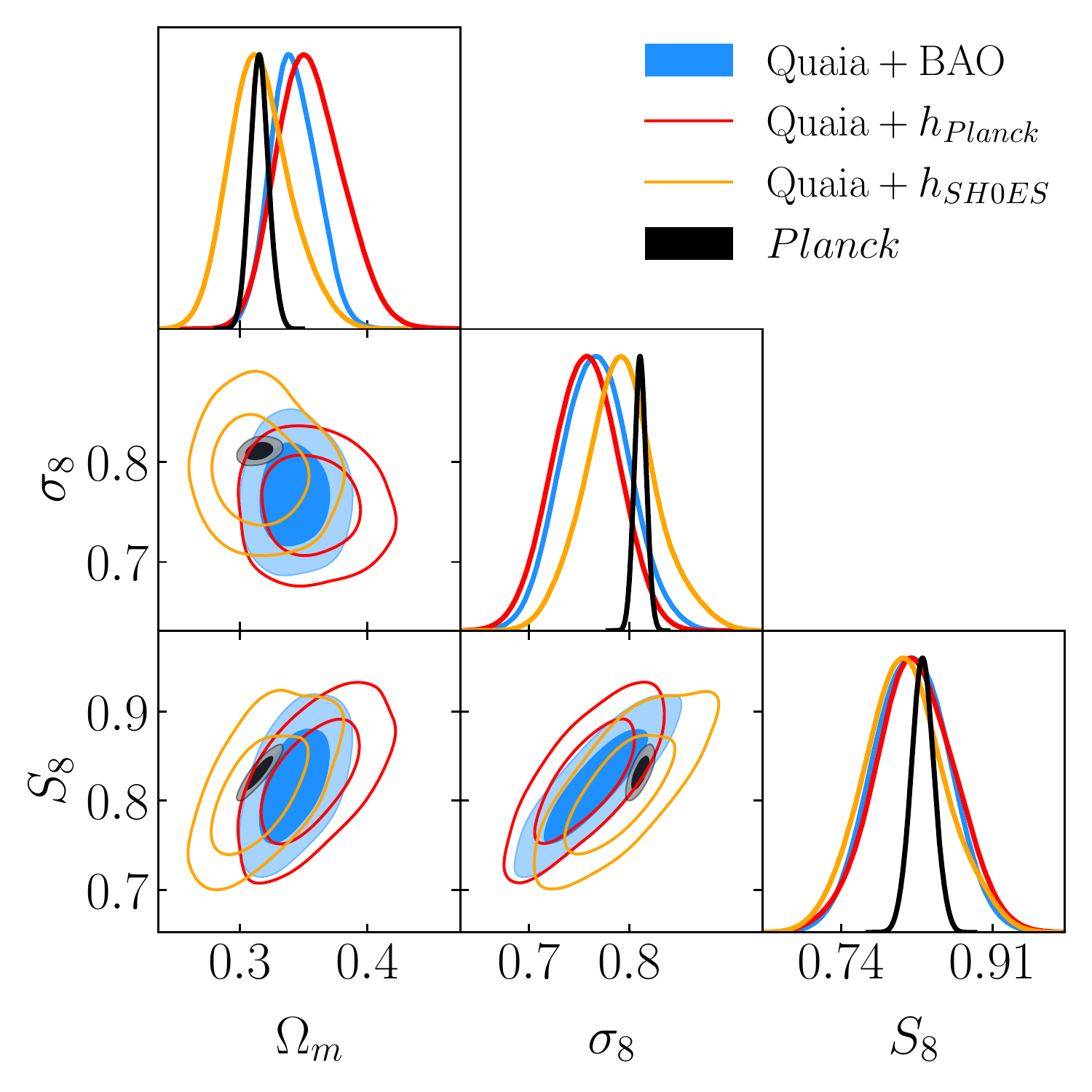}
       \caption{Constraints on $(\sigma_8,\Omega_m,S_8)$ from our analysis using a BAO prior (our fiducial case, in blue), and imposing a prior on the expansion rate today, from \planck (red) and from the SH0ES collaboration (orange). Although the constraints on $\sigma_8$ are only mildly affected by the choice of prior, the effect on $\Omega_m$ is larger, shifting it by up to $1.3\sigma$.}\label{fig:triangle_priors}
     \end{figure}
     First, we consider the option of adding, to our fiducial BAO prior, a constraint from the angular sound horizon scale at recombination, $\theta_{\rm LSS}$, from the positions of the peaks in the CMB power spectrum. This is one of the cleanest CMB observables, purely geometrical by nature, and is remarkably well measured ($\theta_{\rm LSS}=1.04109\pm0.00030$ \cite{1807.06209}). Following \cite{2105.03421}, we include this constraint by imposing a Gaussian prior on the combination $\Omega_mh^3 = 0.09633 \pm 0.00029$ (and varying this quantity instead of $h$ in our MCMC chains). The result is shown in row 17 of Table \ref{tab:results} and Fig. \ref{fig:results_tests}. The addition of this prior leads to a significant tightening of the constraint on $\Omega_m$, which now agrees fully with \planck with almost matching uncertainties. The constraint on $\sigma_8$, on the other hand, remains almost unchanged, with only a $0.3\sigma$ downwards shift (away from \planck), and a $5\%$ tightening of its uncertainty. It is worth noting that, in spite of the relatively large change in the posterior distribution of $\Omega_m$, the best-fit model in this case is still a good fit to the data (the $\chi^2$ only increases by $\sim$2 with respect to our fiducial best fit).

     We then study the impact of replacing the BAO prior by a Gaussian prior directly on $h$. We consider two possibilities: a \planck prior of the form $h=0.674\pm0.005$, and a supernova prior from the SH0ES team \citep{2112.04510}: $h=0.7304\pm0.0104$. The results are shown in rows 18 and 19 of Table \ref{tab:results} and Fig. \ref{fig:results_tests}. The 2D constraints in the $(\sigma_8,\Omega_m,S_8)$ plane are additionally shown in Fig. \ref{fig:triangle_priors}. Using the \planck prior does not lead to significant changes in the final constraints beyond a broadening of the $\Omega_m$ uncertainty. Interestingly, using the SH0ES prior instead, shifts $\Omega_m$ and $\sigma_8$ closer to the values preferred by \planck (by $\sim1.3\sigma$ in the case of $\Omega_m$). The origin of the $H_0$ tension thus has an impact on the interpretation of our constraints (as would be the case for any data set displaying a degeneracy with $h$).

\section{Conclusions}\label{sec:conc}
  In this work, we have presented a study of the tomographic clustering of quasars in the \Quaia sample, a quasar catalog selected from the third \Gaia data release in combination with \unWISE \cite{quaia}. The analysis is based on the angular auto-power spectrum of \Quaia in two redshift bins, centered at $z=1.0$ and $z=2.1$, as well as the cross-correlation with CMB lensing data from \planck.

  We find that, thanks to the large range of scales made available by \Quaia's large cosmological volume, we are able to break the degeneracy between $\Omega_m$ and $\sigma_8$ that is usually present in tomographic large-scale structure analyses with lower-redshift samples. We are able to constrain $\sigma_8$ at a level that is competitive with (and in some cases better than) the precision obtained by state-of-the-art weak lensing and galaxy clustering experiments (see Fig. \ref{fig:results_experiments}). Furthermore, our measurement is highly complementary to those datasets, since it makes use of significantly higher-redshift data, and mostly linear scales, where the modelling of astrophysical systematics is, arguably, cleaner. Our constraints on the amplitude of matter fluctuations, $\sigma_8=0.766\pm0.034$, and on the fractional abundance of non-relativistic matter, $\Omega_m=0.343^{+0.017}_{-0.019}$, are compatible with \planck at the $1.4\sigma$ level. We also find no evidence of the so-called ``$S_8$ tension'', although our data is less sensitive to that parameter combination than weak lensing experiments. We have shown that our measurement is robust against a large suite of potential systematic effects and analysis choices, including the assumed scale cuts, the assumed redshift evolution of the quasar bias, the impact of non-Poissonian shot noise, uncertainties in the redshift distribution of our samples, contamination from residual sky systematics in the galaxy overdensity map, a mis-estimation of the magnification bias slope, and the choice of priors on other cosmological parameters. Furthermore, we find that our measurements of the quasar bias is in reasonable agreement with previous studies. 
  
  When extracting constraints from each of the two redshift bins of the catalog independently, we find that the higher redshift sample favours a value of $\sigma_8$ that is lower than the \planck measurement at the $\sim$2.5$\sigma$ level. This is caused by the relatively low amplitude of the quasar-lensing cross-correlation in that bin, and drives the lower value of $\sigma_8$ found in our fiducial analysis. This result is robust against all sources of uncertainty in the modelling of the \Quaia sample we have explored (quasar bias evolution, redshift distribution calibration, magnification bias). The explanation for this low amplitude thus most likely lies in the CMB lensing map itself. In particular, by reanalysing the data using an alternative CMB $\kappa$ map, constructed using only polarization information, we have found compelling (although not definitive) evidence of potential contamination from extragalactic foregrounds, which would lower the amplitude of this particular cross-correlation. The most likely foreground source responsible for this effect would be the Cosmic Infrared Background, given its relevance at $z\gtrsim2$. We find that, using the polarization-only $\kappa$ map, the amplitude of the quasar-$\kappa$ cross-correlation can grow by up to $\sim$20--30\%, although the significantly higher noise level of this map prevents us from detecting this shift at more than $\sim$2$\sigma$. This correction is substantially larger than the nominal level of contamination due to CIB leakage in the CMB lensing power spectrum determined by \planck (at the level of $\sim$1\%), but it is able to bring our results into complete agreement with the \planck constraints.

  Fully understanding the presence and origin of this potential contamination would be highly relevant, since it could have a substantial effect on other cross-correlation studies involving high-redshift datasets (such as \cite{1712.02738,1903.07046,1904.13378,2011.01234,2305.07650}). A better characterisation of the effect could be achieved using more sensitive CMB lensing estimators that are still robust against CIB contamination, such as including $TE$ information, or using hardening methods \cite{2007.04325}. The use of high-resolution ground-based CMB lensing data \cite{2004.01139,2304.05203} will also be able to shed light on this effect in the near future.

  Another caveat of our analysis is the degeneracy with the value of the Hubble constant $h$, which we mitigated by making use of low-redshift BAO data. In the future, other spectroscopic quasar catalogs, such as the sample targeted by DESI \cite{2208.08511}, may be able to break this degeneracy independently, making use of their own measurements of the redshift-distance relation. We also showed that our final constraints, especially in the case of $\Omega_m$, are sensitive to the choice of prior on $h$ (if one is used), and thus may be affected by the ongoing tension between high redshift CMB and large-scale structure data, and direct measurements from supernovae.

  Our analysis has demonstrated the significant and complementary constraining power of high-redshift large-scale structure data sets covering large volumes. The characteristics of the \Quaia catalog (comoving volume, redshift precision, homogeneity, reduced observational systematic effects enabled by space-quality data), make it ideal to carry out a number of other cosmological studies. These include constraining the level of large-scale homogeneity \cite{quaia-homogeneity} and isotropy \cite{quaia-isotropy}, and detecting ultra-large scale effects, such as primordial non-Gaussianity \cite{quaia-fnl}. The main limiting factor of this sample is its relatively low density, which significantly reduces the amount of information that can be extracted from small (but still only mildly non-linear) scales. This will likely be improved upon by future quasar catalogs from next-generation experiments such as DESI \cite{2208.08511} or the Vera C. Rubin Observatory \cite{2017IAUS..324..330I}, which will be able to improve on the results obtained here. Future \Gaia data releases may also enable improvements to the catalog in terms of redshift precision, quasar identification, and decontamination, which could allow for improved measurements.

\section*{Data access and software}
  The per-bin selection functions used in this analysis, and the complete set of power spectra, covariance, and associated metadata used in our fiducial analysis are publicly available at \url{https://doi.org/10.5281/zenodo.8098635}. The public \Quaia catalog can be found at \url{https://doi.org/10.5281/zenodo.8060755}.

  This work made extensive use of Astropy \cite{the_astropy_collaboration_astropy_2022,the_astropy_collaboration_astropy_2018,the_astropy_collaboration_astropy_2013}, cobaya \cite{2005.05290}, healpy \cite{gorski_healpix_2005,zonca_healpy_2019}, matplotlib \cite{Hunter2007}, NaMaster \cite{1809.09603}, numpy \cite{VanDerWalt2011}, and scipy \cite{Virtanen2020}.

\begin{acknowledgments}
  We would like to thank Nestor Arsenov, Michael Blanton, Julien Carron, Gerrit Farren, Pedro Ferreira, Andras Kovacs, Anthony Lewis, Blake Sherwin, Lyuba Slavcheva-Mihova, An\v ze Slosar, and Abby Williams for useful comments and discussions. DA acknowledges support from the Beecroft Trust, and from the John O'Connor Research Fund, at St. Peter's College, Oxford.
  GF acknowledges the support of the European Research Council under the Marie Sk\l{}odowska Curie actions through the Individual Global Fellowship No.~892401 PiCOGAMBAS and of the Simons Foundation.
  K.S.F. is supported by the NASA FINESST program under award number 80NSSC20K1545. 
  CGG acknowledges support from the European Research Council Grant No:  693024 and the Beecroft Trust. 
  We made extensive use of computational resources at the University of Oxford Department of Physics, funded by the John Fell Oxford University Press Research Fund, and at the Flatiron Institute.
\end{acknowledgments}

\appendix
\section{Simulation-based assessment of CMB lensing contamination}\label{app:kappa_pol}
   To further investigate the potential evidence of contamination described in Section \ref{sssec:res.rob.kappa_pol}, we performed a consistency test of our finding by cross-correlating the \Quaia sample with different CMB lensing maps of the PR3 release. For the PR3 release, in fact, we have access to a complete set of paired signal only ($\kappa^{s}$) and reconstructed $\kappa$ maps obtained running MV, polarization-only or temperature-only lensing reconstruction ($\kappa^{s}_{\rm MV}, \kappa^{s}_{\rm Pol}, \kappa^{s}_{\rm TT}$ respectively) on simulated CMB maps that contain a full end-to-end treatment of the whole Planck data processing chain \cite{ffp10}. As such, those simulated reconstructed $\kappa$ maps represent statistical realizations of our best knowledge of the Planck lensing data (including instrumental systematics, realistic correlated noise, and foreground residuals after component separation) that can be used to model more accurately the expected difference in bandpowers amplitude in the cross-correlation between \Quaia quasars and different versions of the CMB lensing map. The PR4 has a similar data set only for the GMV and MV lensing reconstruction and, as such, we could only perform consistency tests using the data-driven null model discussed in Section \ref{sssec:res.rob.kappa_pol}. Our pipeline operated as follows:
   \begin{enumerate}
     \item For each $\kappa^{s}$ realization, we generated a constrained correlated Gaussian realization of the clustered component of the \Quaia sample. For this we assumed the theoretical correlation between the \Quaia quasars and CMB lensing described in Section \ref{sec:th}, for the same cosmology used to obtain the $\kappa^{s}$ realizations.
     \item For each of these signal-only realizations of the \Quaia quasar overdensity, we added a shot-noise component modeled with random realizations of the catalog\footnote{More details on these simulations are given in \cite{quaia-fnl}.} to produce realistic \Quaia-like quasar overdensity maps $g^{s}_{\rm\Quaia}$. 
     \item For each reconstructed CMB lensing map and its corresponding correlated $g^{s}_{\rm\Quaia}$ map, we computed $C_\ell^{\kappa^{s}_{\rm MV}g^{s}_{\rm\Quaia}}$, $C_\ell^{\kappa^{s}_{\rm Pol}g^{s}_{\rm\Quaia}}$ and their difference for all the different CMB lensing reconstruction methods\footnote{We corrected each estimate of cross-correlation power spectra by the corresponding lensing normalization correction for PR3 following the approach used in Section \ref{ssec:meth.kappa}.}.
     \item We evaluated the covariance of the shifts in bandpowers ${\rm Cov}(C_\ell^{\kappa^{s}_{\rm MV}g^{s}_{\rm\Quaia}}-C_\ell^{\kappa^{s}_{\rm Pol}g^{s}_{\rm\Quaia}})$ averaging over all available realizations.
     \item Using this covariance, we computed the $\chi^2$ for each simulation realization as well as its distribution. We repeated the same procedure for the $\chi^2$ distribution of each bandpower.
   \end{enumerate}
   We then assessed the significance of the observed shifts in the data bandpowers $C_\ell^{g\kappa_{\rm MV}}-C_\ell^{g\kappa_{\rm Pol}}$ comparing the value obtained for the $\chi^2$ performed on data with the $\chi^2$ distribution obtained from simulations using only the multipoles in the $30\leq\ell\leq 600$ range. We adopted the same approach for the $\chi^2$ test for each bandpower observed for the data. We found a PTE of 6\% for the full spectrum, supporting the conclusions derived with our data-based model for PR4, although with worse PTE approaching a ~$2\sigma$-level tension as we found fitting the $A_{\rm pol}$ parameter. When considering the $\chi^2$ test per bandpower, we found that almost all the observed shifts in the data are consistent with simulations (PTEs above 5\%). However, the bin at $\ell\approx 100$ and the one at $\ell\approx 370$ have a PTE of about $2\%$, hinting to some mild evidence of inconsistency with the hypothesis of bandpower shifts only due to statistical fluctuations. The feature at $\ell\approx 100$ in particular is the one driving the low (yet acceptable) PTE for the full $\chi^2$ test. We also note that lowering the covariance estimated from our Monte Carlo approach by a factor of 20\% to mimic the gain in noise obtained in PR4 GMV analysis compared to PR3 did not change these conclusions. We also repeated a similar test comparing the MV with TT-only reconstruction, the MV with the TT lensing reconstruction obtained explicitly nulling the tSZ contribution when performing component separation, and the TT with the TT-only tSZ-deprojected reconstruction. In all these cases we found that the observed shifts in the $g\kappa$ bandpowers are consistent with the differences induced by statistical fluctuations, reinforcing the conclusion that, among the extragalactic foregrounds, the CIB emission rather than tSZ is more likely to drive the low amplitude of the $g\kappa$ signal at high redshift.

\bibliography{gaia_qsoXcorr,non_ads}{}

\end{document}